\documentclass[aps,prc,twocolumn,superscriptaddress]{revtex4-2}

\usepackage{graphicx}% Include figure files
\usepackage{dcolumn}% Align table columns on decimal point
\usepackage{bm}% bold math
\usepackage{longtable}
 \usepackage{threeparttable}
 
\begin{document}

\title{On fluctuation properties of MACS}

\author{M.~Krti\v{c}ka} %https://orcid.org/0000-0002-0083-8327
\affiliation{Faculty of Mathematics and Physics, Charles University, Prague, 180 00, Czech Republic}
\author{A.~Couture} %https://orcid.org/0000-0002-0861-3616
\affiliation{Los Alamos National Laboratory, Los Alamos, New Mexico 87545, USA}
\date{\today}

\begin{abstract}
%The Maxwellian Average Cross Section is usually predicted with help of the statistical codes that do not take into account fluctuations of individual resonance sequences. In reality, for nuclei with $s$-wave resonance spacing comparable to MACS temperature $T$, the expected deviations of actual cross sections from the expectation value are substantial. In this contribution we investigate the size of these fluctuations and contribution of various effects to them. {\bf Should we add anything? MK supposes that the abstract will be rephrased at the end.}
\begin{description}
\item[Background] The Maxwellian Average Cross Section (MACS) is usually calculated with help of the statistical codes that do not take into account fluctuations of individual resonance parameters. The actual MACS can substantially deviate from its expectation value.
\item[Purpose] This work focuses on description of various sources and aspects of these fluctuations.
\item[Methods] Simulated resonance sequences considering all the fluctuations involved in the statistical model are used for this purpose. Contribution of various sources of the fluctuations and neutrons with different orbital momenta are thoroughly investigated and described. Impact of available cross section data at low neutron energies is also checked.
\item[Results] 
Based on this analysis simple empirical formulae for estimating relative fluctuations of MACS using solely $s$-wave resonance spacing $D_0$ and the temperature of the stellar environment are derived. It is shown that real nuclei follow the proposed formula well.
\item[Conclusions] 
The expected MACS fluctuations increase with $D_0$ and can be significant. While for isotopes with $D_0$ in eV range the possible deviation from calculated values are small, for nuclei with $D_0$ in the keV range the full-width half-maximum of the expected distribution is larger than about 25\% and 10\% for astrophysically relevant temperatures of $kT=8$ and 30 keV. At least for these nuclei the predictions from statistical codes must be taken with caution.
\end{description}
\end{abstract}

\maketitle

\section{Introduction}

%{\it This is the very first version of MK. I do not know much about this:-) References are needed. I am not sure about relevant ones.} {\bf An update is needed.}

Neutron-induced reactions play an important role in a wide range of scientific and technical disciplines, impacting such diverse arenas as the origins of the elements, understanding the history of the solar system, power generation from nuclear reactors, nuclear forensics, and stockpile stewardship \cite{BBF57,WIP97,FiW23,HGH20}. Radiative neutron capture, the process by which a neutron is absorbed by the nucleus creating a nucleus of the same element, but one mass unit heavier, is particularly interesting as it is energetically allowed for almost all bound nuclei.  Further, since the neutron is uncharged,
there is no Coulomb barrier inhibiting reactions at low energies, unlike in most charged particle-induced reactions.  As a result, understanding neutron capture cross sections is essential across the chart of nuclei. As one relevant example, the reaction is responsible for the cosmic production of almost all stable nuclei above iron.

Our experimental understanding of neutron capture cross sections is driven primarily by two measurement techniques -- activation and time-of-flight measurements \cite{ABC21}.  Activation provides an integral cross section over a known neutron flux, often with very high precision.  The technique is largely limited to samples which produce an unstable reaction product with a convenient lifetime for post-irradiation analysis, however, very small, impure samples can often be investigated.  Time-of-Flight (TOF) measurements provide energy-differential cross sections, often over a very wide neutron energy range, albeit at a somewhat reduced precision. Further, since TOF measurements rely on the detection of prompt photons, there are fewer limitations on accessible isotopes than for activation. Significantly larger samples are required, and the necessary isotopic enrichment depends on the details of both the nucleus and the detection scheme.  Finally, both the lifetimes of the samples and sample activity typically limit both techniques to measurements on isotopes at most one mass unit from stability.

For almost all applications requiring neutron capture cross sections, the environments are such that thermalization timescales are much shorter than the nuclear reaction timescales, even in such rapidly evolving scenarios as an explosive \emph{r}-process \cite{WIP97}.  As a result, these environments are all in local thermal equilibrium.  The motion of both the target atoms as well as the neutrons is determined not by the energy at which they are created, but rather by the temperature of the environment. A Maxwellian velocity distribution defines the motion of the particles and the neutron-capture cross section is needed not for a fixed neutron energy, but rather weighted over the range of energies corresponding to that temperature. This weighted cross section is referred to as the Maxwellian-Averaged Cross Section (MACS). Critically, the energy range needed to determine the appropriate MACS depends on the temperature of the environment $T$ and is given in energy by $kT$.  Cross sections, and thus measurements, from neutron energy $E_{n} \approx 0.02-6\, kT$ typically contribute significantly to the MACS. Stellar reaction rates are at least needed over energies ranging from $kT=5-100$~keV, which defines a range over which neutron capture cross sections need to be known.

While there exist some cross section measurements on almost all stable isotopes, in many cases the energy range of those measurements does not cover the full energy range needed for reliable calculation of the corresponding MACS.  Further, for unstable isotopes, there are very few measurements to date. To bridge these gaps, we rely on theoretical tools. In particular, MACS are typically obtained from calculations made within the Hauser-Feshbach (HF) formalism as the compound-nucleus reaction often dominates the radiative neutron capture. This formalism uses average resonance quantities which give a definite value of MACS \cite{TALYS,CoH}.

In the limit of neutron capture being determined by compound nuclear reactions, which is true for most heavy nuclei near stability, where experimental data exist, the MACS is an integration over many individual resonances, which gives a single, unique value for the MACS. That same series of resonances can be used to determine what the \emph{average} resonance properties are. An HF calculation would use those average properties to determine the cross section.  However, many different series of resonances can give rise to the same average properties.  It is not \emph{a priori} given that the cross section determined from integration over different series of resonances which give the same average properties result in the same MACS integral.  As a result, the MACS determined from HF calculations carries an inherent uncertainty coming from the uncertainty in how well the average values of the resonance parameters characterize the range of series of resonances that could give rise to them.  

We will call this variation arising from different possible resonance series consistent with the average properties a fluctuation in the MACS.  For cases where the cross section has been directly measured over the entire cross section range of interest, this fluctuation will be zero, but for cases where the cross section is determined from either solely the average properties or from a combination of direct measurement and calculations, there is the possibility of additional uncertainty arising from such fluctuations. The impact of these fluctuations has been partly illustrated in \cite{Rochman17}.

In this contribution we investigate in detail the fluctuations of MACS and their dependence on various underlying physical quantities. We show that the actual values of $\sigma_{MACS}$ may be expected to substantially deviate from the expected values especially for nuclei with $s$-wave (neutrons with orbital momentum $\ell=0$) resonance spacing higher than several hundreds of eV. The only possibility to obtain the cross section with a high precision in these cases is thus to measure its energy dependence, at least for a range of neutron energies or determine it with the activation technique. Impact of available data at low neutron energies on the fluctuation of MACS is also investigated.

In principle, the analysis performed in this contribution is not restricted to MACS but can be made for the neutron cross section in general. However, the exact quantification of the fluctuations then depend on averaging interval, while the MACS is constrained by a well-defined neutron flux profile and it is a quantity of high practical importance. 

%%%%%%%%%%%%%%%%%%%%%%%%%%%%%%%%%%%%%%%%%%%%%%%%%%%%%%%%%%%%%%%%%%%%%%%%%%%%%%%

\section{\label{sec:basic} Basic considerations and generation of resonance sequences}
%{\it Division to different section/subsections can be very different. (Sub)Section titles might be very different. Any suggestion welcome. I am not very happy the titles now:-)}

The MACS is defined as \cite{Beer92}
\begin{eqnarray}
\sigma_{\rm MACS} & = & \frac{\langle \sigma v \rangle}{v_T}
 =  \frac{\int\limits_0^\infty \sigma(v) v \Phi(v) dv}{v_T}  \nonumber \\
& = & \frac{2}{\sqrt{\pi} (kT)^2} \int\limits_0^\infty \sigma(E_n) E_n \exp(-E_n/kT) dE_n, 
\label{eq:macs_beer5}
\end{eqnarray} 
where $v$ is the neutron velocity in the center-of-mass system, $v_T$ is the average neutron velocity, $\Phi(v)$ is the neutron flux, $E_n$ neutron energy, $\sigma(E_n)$ (and $\sigma(v)$) the neutron capture reaction cross section, and $kT$ is the temperature of the environment expressed in energy (as a product of Boltzman constant $k$ and temperature $T$). 

While multiple reaction mechanisms can contribute to the MACS, compound nuclear reactions typically dominate the cross section for all but the lightest nuclei or nuclei near neutron drip-lines.  The compound nuclear reaction mechanism is often treated within the Hauser-Feshbach (HF) formalism adopting nuclear level density and strength functions for neutrons and photons that are used to calculate average decay widths and/or transmission coefficients. 
%{\it Should we use $kT$ or only $T$? Should we use $E_n$ or only $E$? Then we should $k_n$ replace with $k$, Boltzman constant must have a different symbol.}
Within the HF formalism, the cross section $\sigma(E_n) \equiv \sigma^{HF}(E_n)$, used in Eq. (\ref{eq:macs_beer5}), is for given $E_n$ calculated as
\begin{equation}
\sigma^{HF}(E_n) = \frac{\pi}{k_n^2} \sum\limits_{J^\pi,\ell} \frac{g_J T_{n}^\ell T_{\gamma}^{J^\pi}} {T_T^{J^\pi}} W_{n\gamma}^{J^\pi}
\label{eq:HFsigma}
\end{equation}
where $T_n^{\ell} = 2\pi \langle \Gamma_n^{\ell} \rangle / D_J = 2\pi \sqrt{E_n} S_\ell P_\ell$ is the neutron transmission coefficient for neutron with orbital momentum $\ell$, $\Gamma_n^{\ell}$ is the corresponding neutron width, $D_{J^{\pi}}$ the average spacing of resonances of spin $J$ and corresponding parity, $S_\ell$ is the neutron strength function for a neutron with orbital momentum $\ell$, and $P_\ell$ is a penetrability factor for such a neutron. Further, $T_\gamma^{J^\pi} = 2\pi \langle \Gamma_\gamma^{J^\pi} \rangle / D_{J^\pi}$ is the $\gamma$-ray transmission coefficient, $\Gamma_\gamma^{J^\pi}$ is the total radiation width, $g_J = (2J+1) / 2 (2J_T+1)$ the (statistical) spin weighting factor for resonance with spin $J$, $J_T$ the spin of the target nucleus, and $k_n$ is the wave number corresponding to a neutron with energy $E_n$.  The widths fluctuation factor 
\begin{equation}
W_{n\gamma} = \left\langle \frac{\Gamma_n \Gamma_\gamma}{\Gamma_T} \right\rangle \Big/ \frac{\langle \Gamma_n \rangle \langle \Gamma_\gamma \rangle}{\langle \Gamma_T \rangle}
\end{equation}
then takes into account the correlation of the partial and total decay widths in individual resonances. Here, $\Gamma_T$ is total width of the resonance given by sum of all decay widths. Symbol $\langle \bullet \rangle$ then indicates the average value of the corresponding quantity.

The presence of individual, isolated resonances makes the cross section strongly dependent on neutron energy. If $k_n$ (or $E_n$) over a resonance can be considered to be a constant (the resonance width is not very high), then $\sigma_{\rm MACS}$ can be obtained as  
\begin{equation}
\sigma_{\rm MACS} = \frac{2}{\sqrt{\pi} (kT)^2} \sum\limits_i E_{ni} \frac{2\pi^2}{k_{ni}^2} \frac{g_J \Gamma_{ni} \Gamma_{\gamma i}}{\Gamma_{Ti}} \exp(-E_{ni}/kT),
\label{eq:macs_areas}
\end{equation}
where the sum goes over all the resonances $i$.
The term 
\begin{equation}
\frac{g_J \Gamma_{ni} \Gamma_{\gamma i}}{\Gamma_{Ti}} =
\frac{1}{2\pi} \int\limits_{-\infty}^{+\infty} \frac{g_J \Gamma_{ni} \Gamma_{\gamma i}}{(E-E_{Ri})^2+\Gamma_{Ti}^2/4} dE \equiv R_k
\label{eq:area}
\end{equation}
is the resonance kernel and corresponds to an integration of the Breit-Wigner form over all neutron energies. Actual parameter values are expected to fluctuate from resonance to resonance. As discussed below, three primary components that are expected to fluctuate in impactful ways include the neutron width $\Gamma_{ni}$, the gamma width $\Gamma_{\gamma i}$, and the neutron resonance energy $E_{Ri}$.  Individual channels contributing to $\Gamma_{n}$ are expected to fluctuate according to the Porter-Thomas distribution. There is also fluctuation of $\Gamma_{\gamma i}$ as well as uncertainty in positions of individual resonances; parameterization of these latter two fluctuations is generally more complicated and uncertain and is discussed below.

If the number of resonances in the energy region relevant for the MACS is high, the resulting uncertainty in predicted cross section due to unknown parameters of individual resonances is small. The fluctuations in MACS are then correspondingly negligible. 

On the other hand, if the resonance density near neutron separation energy $S_n$ is relatively low, one can expect that the fluctuations of individual resonance parameters can lead to significantly different results for a particular realization of a resonance sequence even if all the underlying average quantities (resonance spacing $D_{J^\pi}$, neutron strength functions for individual orbital momenta $S_l$, average radiative width $\langle \Gamma_\gamma \rangle$) are well known. 

%%%%%%%%%%%%%%%%%%%%%%%%%%%%%%%%%%%%%%%%%%%%%%%%%%%%%%%%%%%%%%%%%

\subsection{Numerical Approach to Resonance Treatment} \label{sec:bc_na}
To check and quantify the expected fluctuations of MACS we generated many random sequences of neutron resonances using aforementioned average resonance parameters and expected fluctuation properties of involved quantities. Namely, two thousand sequences were generated unless $D_0$ was on the order of eV, in which case $500-1000$ sequences were used.  For each generated sequence we then calculated $\sigma_{\rm MACS}$ exploiting Eq.~(\ref{eq:macs_areas}). 
The adopted approach allows us to determine separately the contribution of neutrons with different orbital momenta $\ell$ and check the impact of fluctuations of different quantities on the MACS.

A few simplifying assumptions were applied while calculating $\sigma_{\rm MACS}$ from individual resonance sequences:
\begin{itemize}
\item 
Resonances were generated only up to neutron energy $E_n=10kT$. The contribution of higher neutron energies to MACS is strongly suppressed due to the shape of the neutron spectrum combined with the fact that the cross section does not significantly increase with $E_n$ even if no other reaction channels (like inelastic neutron scattering) are opened. Specifically, the cumulative contribution to the MACS reaches $\approx 99.9\%$ for $E_n < 8 kT$ and the fraction of MACS originating from $E_n > 10 kT$ reaches at maximum only about $10^{-4}$~\cite{Beer92}. 
\item 
$S_l$ and $\langle \Gamma_\gamma \rangle$ were kept constant over the whole region of considered $E_n$. While this does not  exactly correspond to the situation in nature (especially for larger $kT$), the influence of the change of these quantities on fluctuations is small. 
\item 
Cases where direct capture contributes significantly to the MACS are not investigated in this work.  Specifically, the contribution of $s$-wave direct capture as well as of negative-energy resonances was neglected. This contribution is at most a few percent (for the lowest relevant $kT$ and significantly decreases with increase of $kT$) and its impact on fluctuations is negligible.
\item 
The constant-temperature (CT) energy dependence of resonance density was followed unless another approach is explicitly noted. The resonance density was thus assumed to increase exponentially with excitation energy as $\rho_{J^\pi}(E = E_n + S_n) = \rho_{J^\pi}(S_n)\cdot\exp(E_n/T_{CT})$ with the temperature $T_{CT}$ taken from Ref.~\cite{Egidy05}. The resonance density was considered to be independent of parity and the spin dependence was given by the spin cut-off parameter $\sigma_c$, if not specified then corresponding to Back-shifted Fermi-gas (BSFG) model from Ref.~\cite{Egidy05}. 
\item 
The channel radius $R$, needed for calculation of penetrability factor for $\ell>0$ neutrons, is adopted in the form of $R=0.123A^{1/3}+0.08$~fm~\cite{Frohner92}.
\item 
Only neutrons with orbital momentum $\ell \leq 3$ were considered. Contribution of higher $\ell$ is typically negligible. 
\end{itemize}

\begin{figure}
\includegraphics[clip,width=0.89\columnwidth]{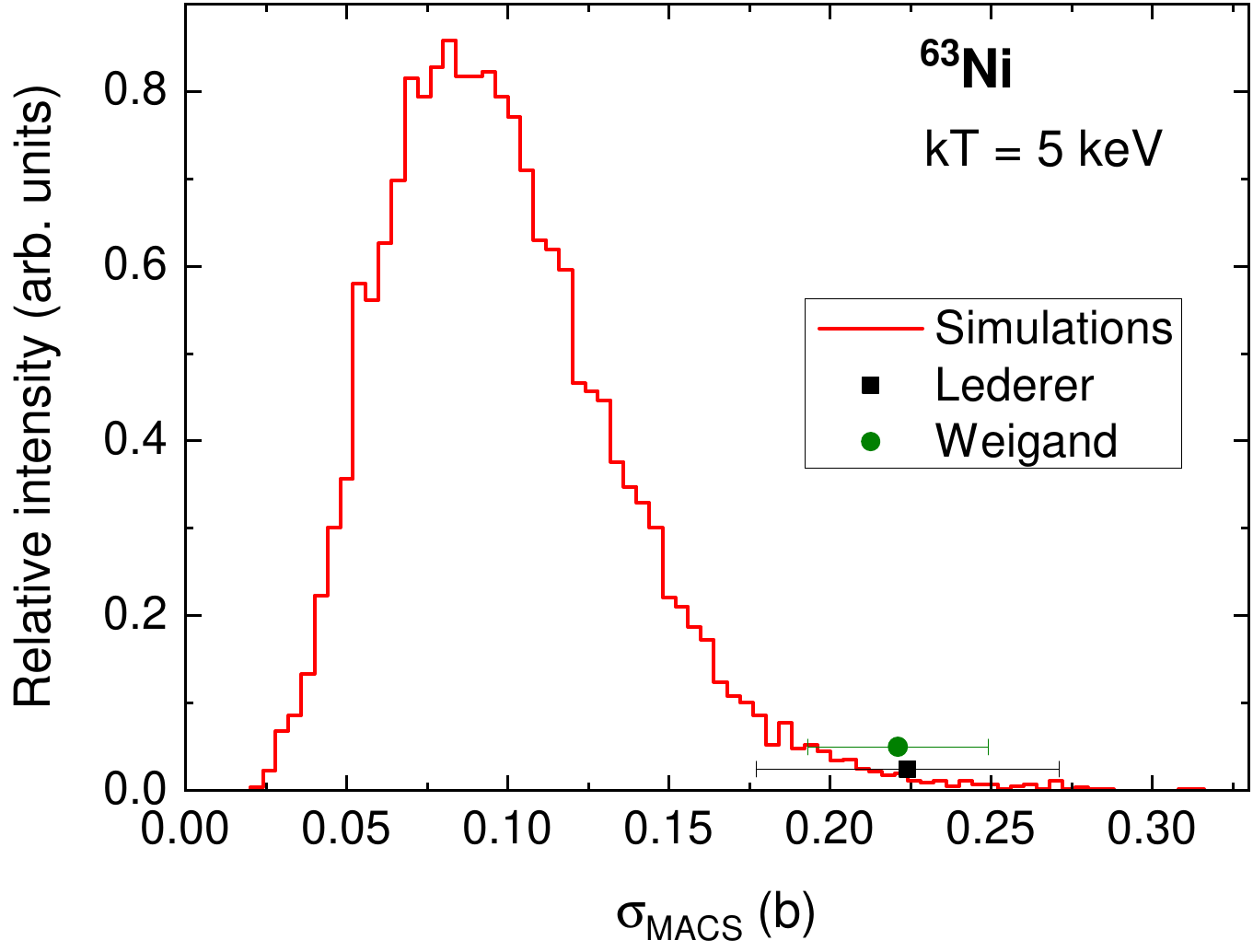}
\caption{Distribution of individual simulated MACS values for $^{63}$Ni target nucleus corresponding to $kT=5$ keV. Experimental values from Lederer {\it et al.} \cite{Lederer13} and Weigand {\it et al.} \cite{Weigand15} are also shown. 
}
\label{fig:macs_distrib}
\end{figure}

\subsection{Adopted Parameters}

The actual distribution of individual MACS values can be asymmetric, as visible from example in Fig. \ref{fig:macs_distrib}. For the description of fluctuation properties in this paper we will dominantly use relative uncertainty $\delta_{\rm MACS}$, given by the ratio of standard deviation of individual simulated MACS values and the average MACS value. 

From nucleus to nucleus, the average resonance parameters vary.  This presents a challenge in attempting to study the general properties of the behavior of the MACS. Thus, to understand behavior of various sources of fluctuations we decided not to check only real nuclei but to perform a systematic study using an {\em artificial nucleus}. This nucleus was based on $^{96}$Zr target. Specifically, the temperature $T_{CT}$ for $^{97}$Zr compound nucleus ($T_{CT}=0.7$~MeV$^{-1}$ \cite{Egidy05}) was adopted and the target spin $J_T$ was considered to be $J_T=0$. The value of $\sigma_c$ from Ref.~\cite{Egidy05} for such an artificial nucleus gives the resonance density ratios at $S_n$: $\rho(J=3/2)/\rho(J=1/2)=1.86$ and $\rho(J=5/2)/\rho(J=1/2)=2.46$. 
%{\it We can transform $\rho$ to spacing. The question is how exactly the spacing should be labeled.}
With this artificial nucleus as a starting point, it is more straightforward to test a range of parameters that real nuclei might actualize. There is no {\em a priori} reason for a choice of the artificial nucleus based on $^{96}$Zr target. Our selection is only motivated by the fact that this nucleus belonged to one of the real tested nuclei that had both lower and higher masses.

The simulations with the artificial nucleus were then performed for three different combinations of $S_\ell$ and $\langle \Gamma_\gamma \rangle$ as a function of $s$-wave ($\ell=0$) resonance spacing $D_0$ that spans several orders of magnitude, from 1 eV up to 30 keV and for $kT=5-100$ keV. This $D_0$ range covers the values found in medium-mass to heavy (stable) nuclei. Different combinations of $S_\ell$ and $\langle \Gamma_\gamma \rangle$ were adopted as the cross section is governed by relation of $\Gamma_n$ (that depends on $S_\ell$) and $\Gamma_\gamma$ when the only decay channels are elastic neutron scattering and gamma emission, see Sec. \ref{sec:widths}. Specifically, we used $S_\ell=0.4 \times 10^{-4}$ in combination with $\langle\Gamma_\gamma\rangle=500$ meV, $S_\ell=6.0 \times 10^{-4}$ with $\langle\Gamma_\gamma\rangle=50$ meV, and $S_\ell=1.5 \times 10^{-4}$ with $\langle\Gamma_\gamma\rangle=200$ meV. The first two combinations represent values that are close to minimum/maximum values found for any stable nucleus (with only a few exceptions for $A>50$ nuclei), the last one is then close to the average over the two extremes. The effect of additional decay channels will be discussed in Section~\ref{sec:art_inel}.

\section{Results and discussion}

\subsection{\label{sec:widths} Impact of different widths on cross section}

As evident from Eq.~(\ref{eq:macs_areas}), if we remove the kinematic factor $E_{ni}\times \exp(-E_{ni})/k_{ni}^2$  (and omit $J^\pi$ and $\ell$ indexes), then in the limit $\langle \Gamma_n \rangle \ll \langle \Gamma_\gamma \rangle$ (corresponding to $R_k\approx g\Gamma_n$ and labeled as limit {\em a} below) the cross section for given $\ell$ is proportional to $g\langle \Gamma_n \rangle /D$. In the opposite limit $\langle \Gamma_n \rangle \gg \langle \Gamma_\gamma \rangle$ ($R_k\approx g\Gamma_\gamma$, labeled as limit {\em b}) it is proportional to $g\langle \Gamma_\gamma \rangle /D$. The radiative neutron capture cross section can thus be governed by both decay widths $\Gamma_n$ and $\Gamma_\gamma$. As their fluctuation properties differ, one should expect different fluctuation behavior of the cross section at different neutron energies. This is clear in limiting cases \emph{a} and \emph{b}. In practice, as shown below the actual cross section behaviour at energies relevant for MACS is very often outside one of these limits which complicates the fluctuation analysis.

%{\bf End of MK version. Replaces also the first paragraph of Sec. III. The text from "Figure 2 indicates ..." (second paragraph) is kept...}

%%%%%%%%%%%%%%%%% Original version - first paragraph of Sec. III %%%%%%%%%%
%{\bf Original version}

%\section{Results and discussion}

%As mentioned in the previous section the radiative neutron capture cross section can be mainly governed by different decay widths $\Gamma_n$ and $\Gamma_\gamma$. As their fluctuation properties are different, one should expect different fluctuation behavior of the cross section or MACS.  This is clear in limiting cases \emph{a} and \emph{b}.  We will develop the behaviour in the intermediate cases below. In practice, as discussed below actual cross section behaviour at relevant energies is very often outside one of these limits which complicates the fluctuation analysis. However, we will describe general fluctuation properties in the rest of the paper and provide {\em effective} formulas for their description.

%%%%%%%%%%%%%%%%% Common version from now %%%%%%%%%%
%{\bf End of original version - common version from now}

\begin{figure}
\includegraphics[clip,width=0.95\columnwidth]{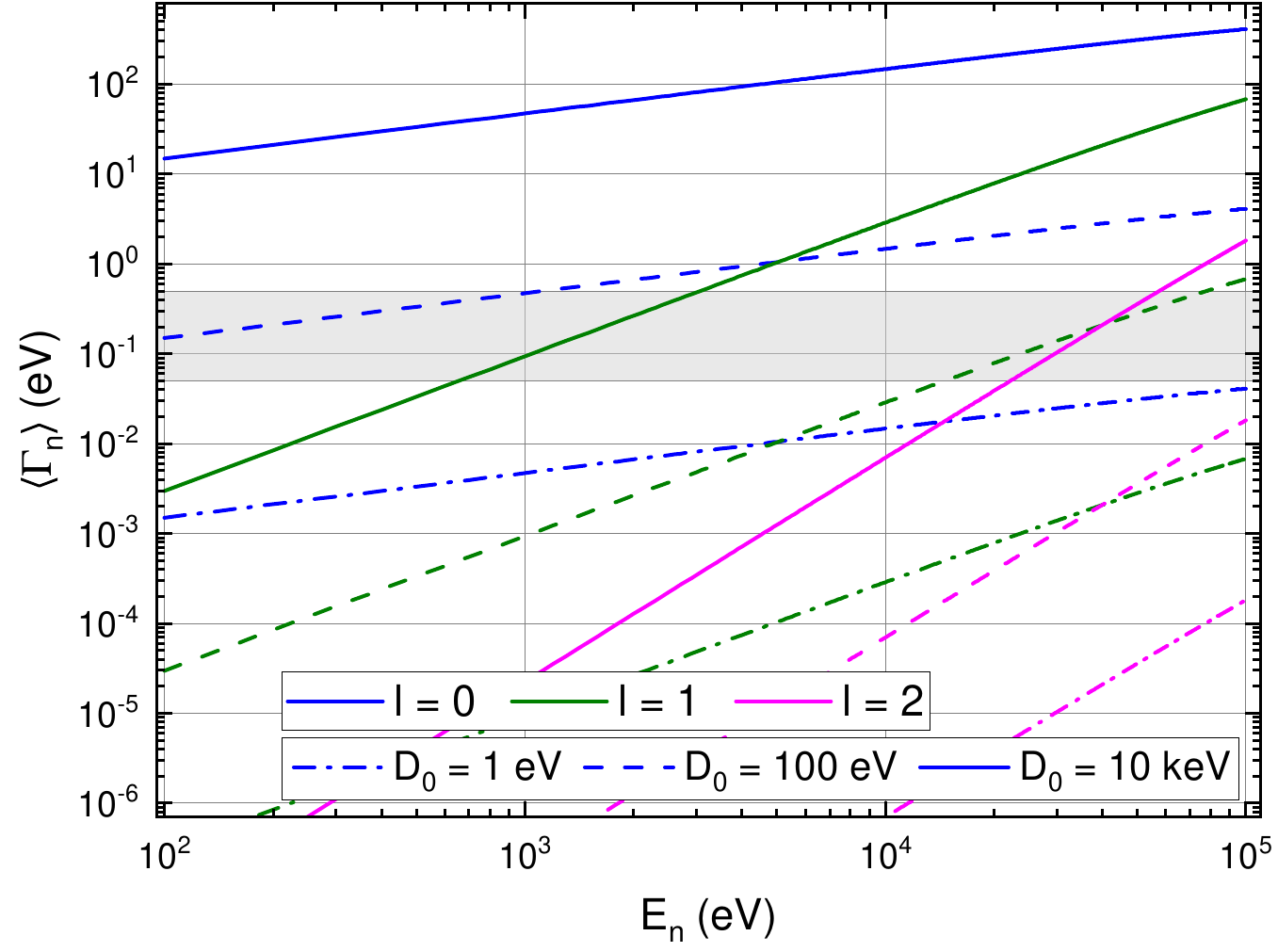}
\caption{Dependence of $\langle\Gamma_n\rangle$ on $E_n$ for three $D_0$ values and $S_\ell=1.5\times 10^{-4}$; artificial nucleus was used in simulations. The horizontal band indicate typical range of $\Gamma_\gamma=50-500$ meV. The line style, which is indicated only for $\ell=0$, holds also for $\ell>0$.
}
\label{fig:gn_dep}
\end{figure}

Figure \ref{fig:gn_dep} indicates values of average neutron widths $\langle \Gamma_n \rangle$ expected in real nuclei. Depending on the actual values of $D_0$ and $S_\ell$, both aforementioned limits $a$ and $b$ might be reached as $\langle \Gamma_\gamma \rangle$ spans values from several tens to several hundreds meV ($50-500$ meV range is indicated by horizontal lines in the figure).   

\begin{figure}
\includegraphics[clip,width=0.95\columnwidth]{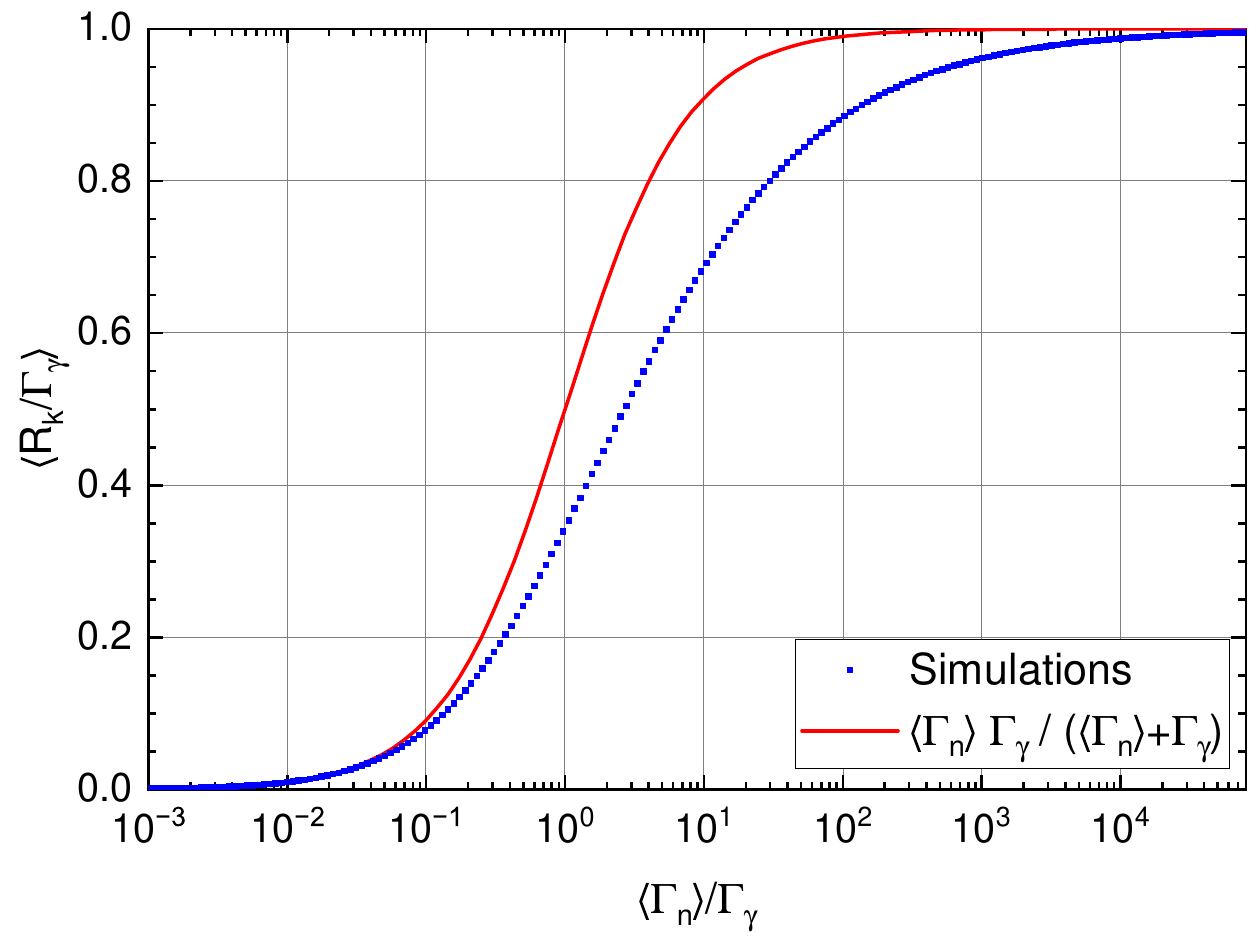}
\caption{Average ratio $\langle R_k/\Gamma_\gamma \rangle$ (assuming $g=1$) as a function of $\langle \Gamma_n \rangle / \Gamma_\gamma$. Red line indicates the dependence if the fluctuation of $\Gamma_n$ is switched off. 
}
\label{fig:k_vs_Gg}
\end{figure}

Neutron widths for one reaction channel at given energy are expected to fluctuate according to a $\chi^2_{\nu=1}$ distribution, known also as Porter-Thomas~\cite{Porter56} distribution, around the expectation value. This fluctuation means that individual values of the resonance capture kernel $R_k$ will significantly fluctuate even for a fixed $\Gamma_\gamma$ and $\langle \Gamma_n \rangle \gg \langle \Gamma_\gamma \rangle$. 
As the cross section at given neutron energy is determined by the sum of $R_k$, the conditions for the validity of both the limits $a$ and $b$ can be checked from the plot of dependence of $\langle R_k /\Gamma_\gamma \rangle$ on the ratio $\langle \Gamma_n \rangle / \Gamma_\gamma$ as plotted in Fig. \ref{fig:k_vs_Gg}. 
The aforementioned limits are reached if $\langle R_k /\Gamma_\gamma \rangle$ are close to zero or one, respectively. Combination of Figs. \ref{fig:gn_dep} and \ref{fig:k_vs_Gg} reveals regions of validity of the limits $a$ or $b$ for given $\ell$. 

For neutrons with $\ell=0$ one is close (within a few percent) to the limit $a$ only for the lowest shown energies (and $D_0$ at maximum of eV); wider energy ranges are close to this limit for $\ell>0$. On the other hand, situation close to the limit $b$ is barely reached even for the largest neutron energies and $D_0$ shown in Fig. \ref{fig:gn_dep} (and only for $\ell=0$). This implies that the cross section contributing to MACS is not close to any of the limits in a substantial range of relevant values (even for a fixed $\ell$).
Note that Fig. \ref{fig:gn_dep} was produced for a specific value of $S_\ell$ while $\langle\Gamma_n\rangle \propto D_0\cdot S_\ell$ (i.e. $\langle \Gamma_n \rangle$ linearly scales with $S_\ell$ and $D_0$). For real nuclei the value of $\langle \Gamma_n \rangle$ values can thus be different.

\begin{figure*}
\includegraphics[clip,width=\textwidth]{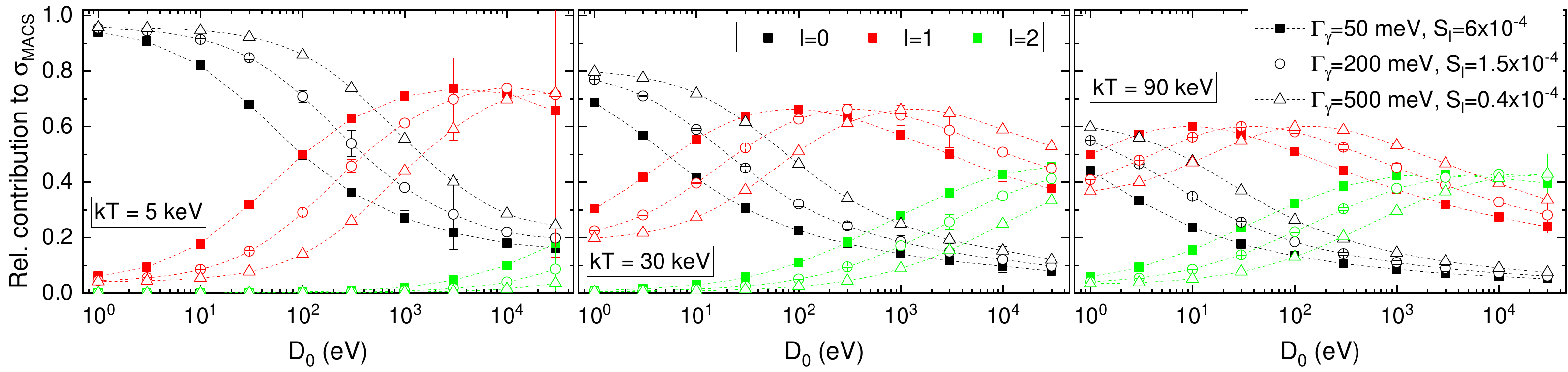}
\caption{
Relative contribution of individual $\ell$ to the cross section for three different $kT=5, 30,$ and 90 keV. An impact of different $\Gamma_\gamma$ and $S_\ell$ (uncertainty shown only for one combination) is indicated. Total assumed cross section included also $\ell=3$ contribution that is not shown (the sum of all curves then can be smaller than one for large $D_0$). Relative fluctuation of $\Gamma_\gamma$ is $\delta_{\Gamma_\gamma}=25\%$. Spacing fluctuated according to Wigner distribution. Symbols correspond to the same $\Gamma_\gamma$ and $S_\ell$ combination for all $\ell$.
}
\label{fig:xs_ell1}
\end{figure*}

In practice, determination of the MACS typically requires consideration of several $\ell$. Information on contribution of neutrons with different $\ell$ to the MACS is thus also of high importance.
Indication of these contributions corresponding to the artificial nucleus can be found in Fig.~\ref{fig:xs_ell1} for three different $kT$ (and three combinations of $\Gamma_\gamma$ and $S_\ell$). The contribution of neutrons with different $\ell$ changes significantly as a function of all involved quantities. The fraction of the cross section at $kT=30$~keV due to neutrons with different $\ell$ for real nuclei is then also indicated in Tab. \ref{tab:parameters}.

The MACS relevant for astrophysical purposes correspond to $kT\approx 5-100$ keV \cite{KGB11}, which means that relevant neutron energy range is a few hundreds eV up to several hundreds keV.  At these energies, hardly any MACS is given by any of the limits $a$ or $b$. The description of the fluctuation properties of MACS for individual isotopes thus likely becomes relatively complicated. It thus seems interesting to check different sources of fluctuations of MACS. This is the subject of analysis in the rest of this contribution.  Following this detailed analysis, we provide {\em effective} formulas for the fluctuation of MACS which can be used to guide uncertainty expectations.

%%%%%%%%%%%%%%%%%%%%%%%%%%%%%%%%%%%%%%%%%%%%%%%%%%%%%%%%%%%%%%%%%
\subsection{Sources of fluctuations \label{sec:sources}}

The three most obvious sources of fluctuations are neutron widths $\Gamma_n$, radiation widths $\Gamma_\gamma$, and individual resonance positions $E_{R}$. The expected fluctuations of $\Gamma_n$ (Porter-Thomas distribution) has already been mentioned earlier in this section.

Radiation widths $\Gamma_\gamma$ show fluctuations whose exact properties might be complicated. We either calculated the distribution using DICEBOX code \cite{Becvar98} or considered that individual $\Gamma_\gamma$ show a fluctuation from a $\chi^2_\nu$ distribution with relatively large $\nu$ (derived from variance of the distribution simulated by DICEBOX). The latter option gives a very reasonable (although not perfect) description of the simulated distribution. The first option (distribution from DICEBOX) was used for simulation of real nuclei, while the second ($\chi^2_\nu$ distribution) for estimates based on simulations of the artificial nucleus. In practice, the $\langle \Gamma_\gamma \rangle$ from DICEBOX simulations do not necessarily exactly reproduce the experimental values and thus all simulated $\Gamma_\gamma$ in the first option were multiplied by a constant factor to get them close to the literature values of $\langle \Gamma_\gamma^{\ell=0} \rangle$ for $s$-wave resonances (as $\langle\Gamma_\gamma\rangle$ for unstable $^{63}$Ni is highly questionable, we used values that are compatible with neighbor odd-$A$ isotopes). The comparison of adopted $\Gamma_\gamma$ values with those from evaluations \cite{Mughabghab18a,Mughabghab18b,RIPL3} can be made from figures given in Table \ref{tab:parameters}.
 %{\it Should we add any figure? Should we say more about NLD/PSFs used? I am not sure? Probably OK.}

Three models were considered for the spacing of resonance energies $E_{R}$.  Individual resonance positions were generated using (i) the spacing of adjacent resonances (nearest-neighbor spacing, NNS) fluctuated according to Poisson (exponential) distribution, (ii) the NNS fluctuated according to Wigner distribution \cite{Wigner59}, similarly to approach applied in Refs. \cite{Rochman13,Rochman17}, and (iii) individual sequences based on predictions of Gaussian Orthogonal Ensemble (GOE) \cite{Dyson63} predicting long-range correlations in resonance positions. Note that the Wigner NNS distribution is almost exactly reproduced by GOE predictions and it is widely believed to be a very good approximation of nature. There are strong indications (at least for medium-weight and heavy nuclei) that the neutron resonance sequences follow very well the GOE predictions \cite{Haq82} but the validity of GOE for description of sequences has also been questioned~\cite{Koehler13}. Despite the fact that the Poissonian NNS is likely not a realistic representation, at least in nuclei with $D_0$ on the order of eV or tens of eV, its simplicity leads to useful insight, see below.

%%%%%%%%%%%%%%%%%%%%%%%%%%%%%%%%%%%%%%%%%%%%%%%%%%%%%%%%%%%%%%%%%
\subsection{\label{sec:effective} Effective number of levels}

At either of the aforementioned limits, $a$ or $b$, the expected fluctuations should follow relatively simple rules. Let us label the number of levels contributing to MACS as $N$. The relative fluctuation $\delta_{\rm MACS}$ then should be: (i) $\delta_{\Gamma_\gamma}/\sqrt{N}$ if the MACS is given solely by $g\Gamma_\gamma$ (in limit $a$) or (ii) $\sqrt{2/N}$, the relative uncertainty of the Porter-Thomas distribution, if the MACS is given solely by $g\Gamma_n$ (in limit $b$). Due to the Maxwellian distribution of neutron energies actual resonances contribute to MACS with different weights. Thus, $N$ should be something like an ``effective number of levels'' $N_{\rm eff}$. We will introduce and discuss $N_{\rm eff}$ as it will provide a very good guide for our discussion on fluctuations.   

We determined $N_{\rm eff}$ in the following way: We fixed both $\Gamma_n$ and $\Gamma_\gamma$ in simulations with an artificial nucleus. We then let the NNS fluctuate according to the Poisson (exponential) distribution. For this distribution the relative uncertainty in the number of levels in an interval should be $1/\sqrt{N_{\rm eff}}$. The relative uncertainty of MACS, $\delta_{\rm MACS}$, deduced from this simulation was thus used for estimating $N_{\rm eff}$ assuming $N_{\rm eff} = (1/\delta_{\rm MACS})^2$. 
We verified that the same $\delta_{\rm MACS}$ (or $N_{\rm eff}$) is obtained if $\Gamma_n$ was not fixed to $\Gamma_n=\langle \Gamma_n \rangle$, but taken from the Porter-Thomas distribution with identical sequence of random numbers. 

\begin{figure}
\includegraphics[clip,width=0.95\columnwidth]{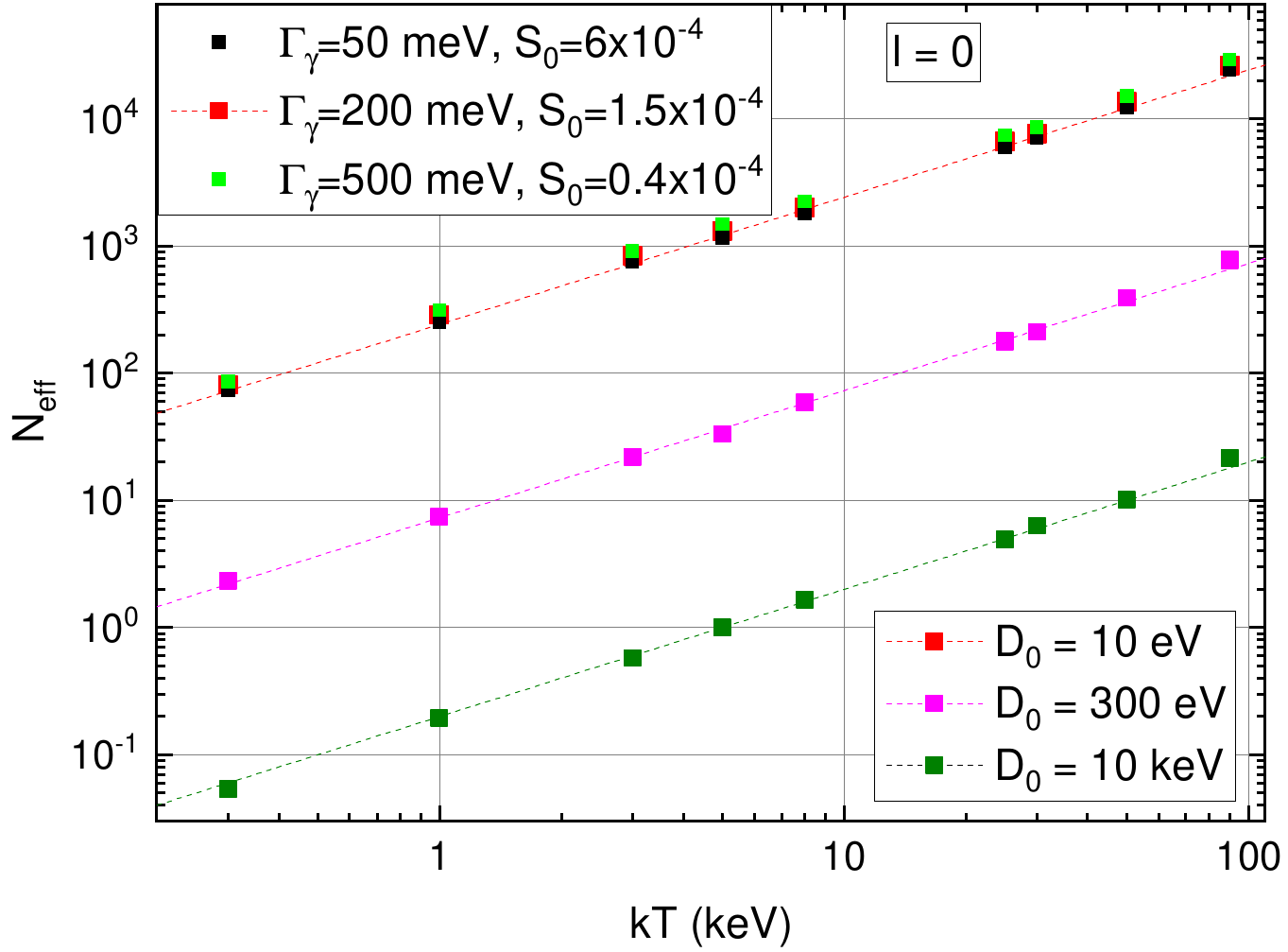}
\caption{Effective number of levels $N_{\rm eff}$, see Sec. \ref{sec:effective}, for $\ell=0$ as a function of $kT$ for artificial nucleus. If not indicated in the legend, the models correspond to $\Gamma_\gamma=200$ meV, $S_0=1.5\times 10^{-4}$. Similar difference of $N_{\rm eff}$ as indicated for $D_0=10$ eV due to different combinations of $\Gamma_\gamma$ and $S_0$ is observed also for other $D_0$. Lines correspond to Eq. (\ref{eq:Neff}). Figures for $\ell=1$ and 2 can be found in Supplemental Material \cite{SupplMat}.
}
\label{fig:effl_kTa}
\end{figure}

\begin{figure}
\includegraphics[clip,width=0.95\columnwidth]{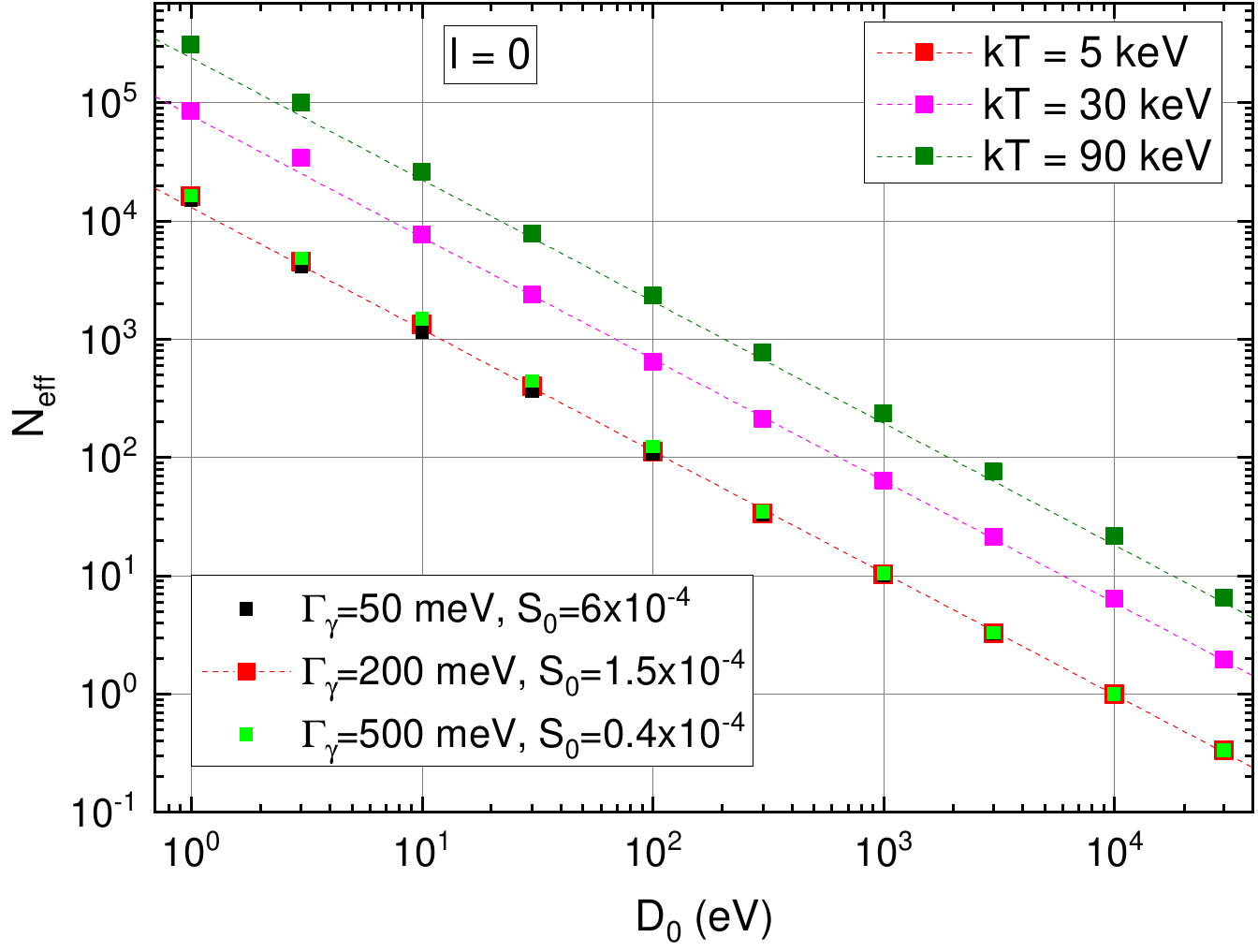}
\caption{Effective number of levels $N_{\rm eff}$ for $\ell=0$ as a function of $D_0$ for three $kT$; artificial nucleus used. If not indicated in the legend, the models correspond to $\Gamma_\gamma=200$ meV, $S_0=1.5\times 10^{-4}$. Similar difference of $N_{\rm eff}$ as indicated for $kT=5$ keV due to different combinations of $\Gamma_\gamma$ and $S_0$ is observed also for other $kT$ values.  Lines correspond to Eq. (\ref{eq:Neff}). Figures for $\ell=1$ and 2 can be found in Supplemental Material \cite{SupplMat}.
}
\label{fig:effl_D0a}
\end{figure}

The dependence of the effective number of levels for $\ell=0$ on $kT$ for a few $D_0$ values can be found in Fig. \ref{fig:effl_kTa}. Figure \ref{fig:effl_D0a} then gives the dependence on $D_0$ for a couple of $kT$. Analogous figures for $\ell=1$ and 2 can be found in  Supplemental Material \cite{SupplMat}. Figure \ref{fig:effl_kTa} shows that for fixed $D_0$ and $\ell=0$, $N_{\rm eff}$ is with a high precision a linear function of $kT$ (independently of $S_\ell$ and $\Gamma_\gamma$ values). 
Although some deviations from this simple dependence are visible, they seem to be small enough to not spoil our goal -- make a simple (empirical) law for practical estimate of expected MACS fluctuations. 
Further, one can expect that $N_{\rm eff}$ for fixed $kT$ is inversely proportional to $D_0$. Actual dependence, see Fig. \ref{fig:effl_D0a} and Supplemental Material, is then slightly steeper than expected, approximately as $D_0^{-1.03}$ for $\ell=0$, as $D_0^{-1.08}$ for $\ell=1$, and as $D_0^{-1.06}$ for $\ell=2$. These deviations from the simple expectation arise from different impact of quantities driving the MACS value for different $D_0$, $\Gamma_n$ (or $S_\ell$) and $\Gamma_\gamma$. A very low impact of different $\Gamma_\gamma$ and $S_\ell$ combinations is visible from Figs.~\ref{fig:effl_kTa} and \ref{fig:effl_D0a}.

Specifically, dependence of $N_{\rm eff}$ on $kT$ and $D_0$ can be well approximated with
\begin{eqnarray}
N_{\rm eff} & = \,\ 2600 \times kT \times D_0^{-1.03} \quad & {\rm for}\ \ell=0, \nonumber \\
N_{\rm eff} & =  14500 \times kT \times D_0^{-1.08} \quad& {\rm for}\ \ell=1,  \label{eq:Neff}\\
N_{\rm eff} & =  32000 \times kT \times D_0^{-1.06} \quad& {\rm for}\ \ell=2, \nonumber 
\end{eqnarray}
where $kT$ is given in keV while $D_0$ is in eV. These dependences are indicated in Figs.~\ref{fig:effl_kTa} and \ref{fig:effl_D0a} and in the corresponding Figures of the Supplemental Material. Unfortunately, such simple dependences work only for individual $\ell$. The combination of more $\ell$ could further modify them as the contribution of different $\ell$ changes, see Fig. \ref{fig:xs_ell1}. Nonetheless, a relatively simple effective description can still work even if all $\ell$ are considered.

\begin{figure}
\includegraphics[clip,width=0.95\columnwidth]{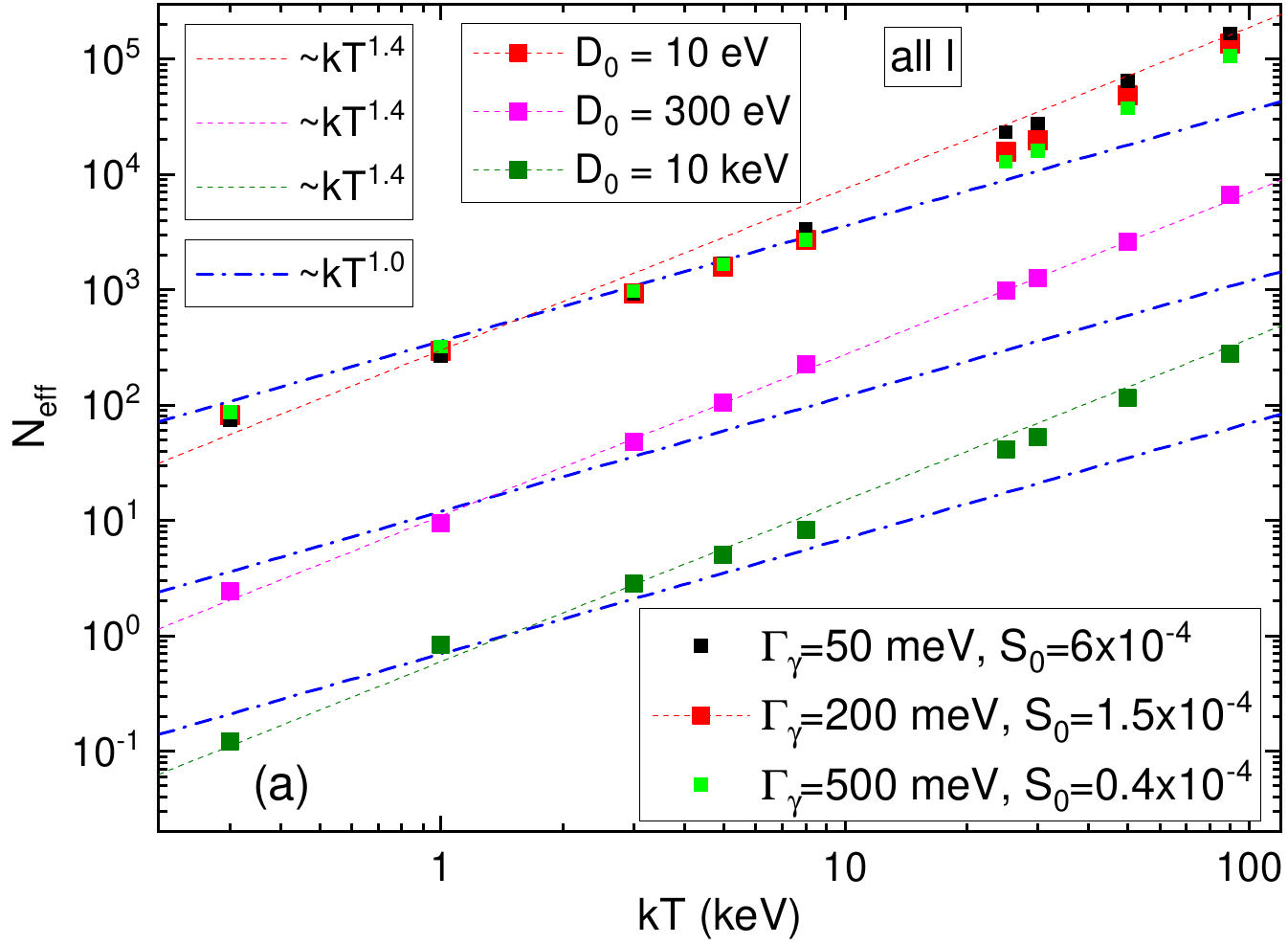}
\includegraphics[clip,width=0.95\columnwidth]{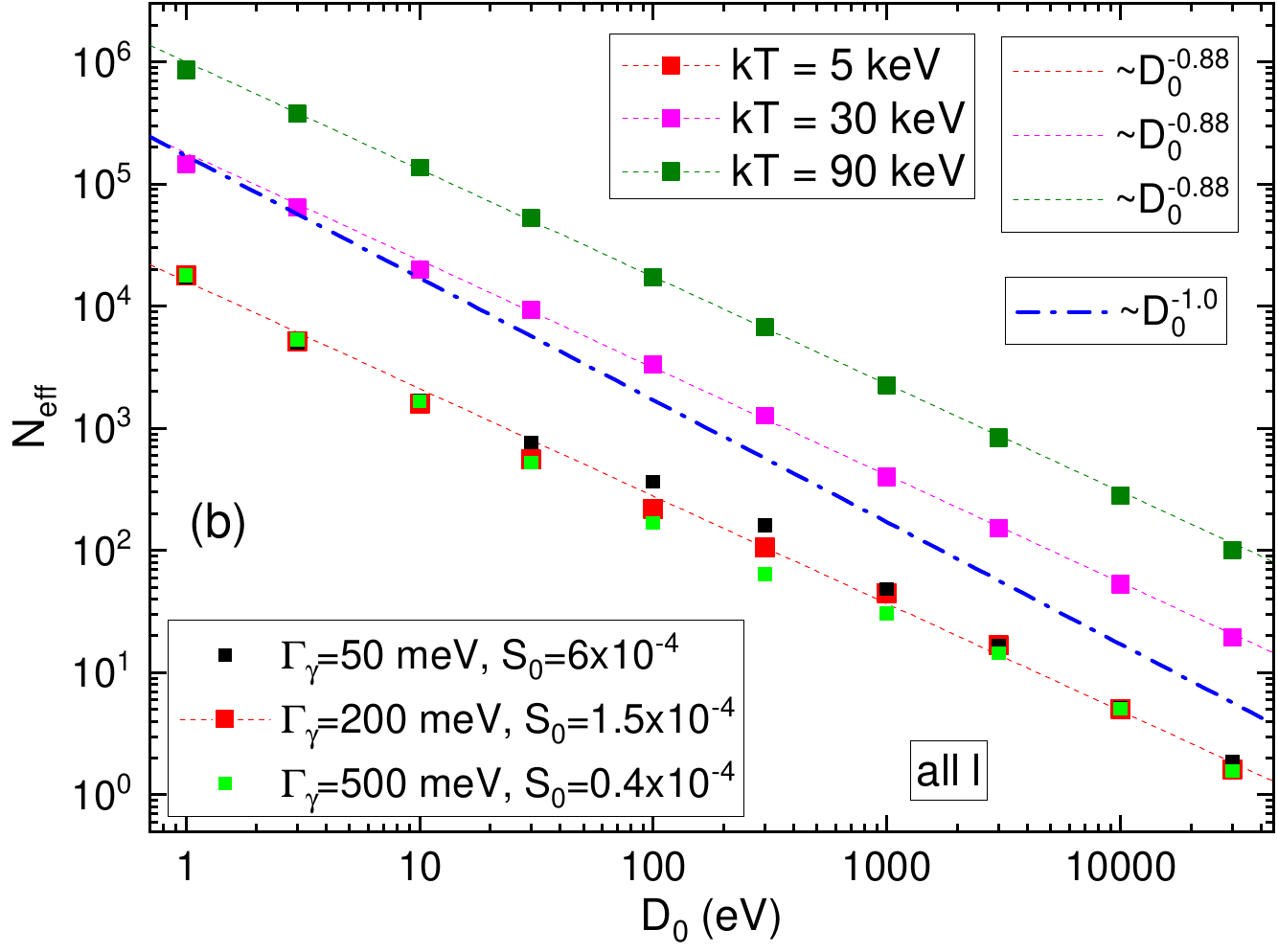}
\caption{Effective number of levels $N_{\rm eff}$ if all $\ell$ are considered as a function of $kT$ (a) and $D_0$ (b). The lines are only to guide the eye and indicate actual dependences.}
\label{fig:effl_tot}
\end{figure}

The combined effect coming from sum of contributions for different $\ell$ then really shows a more complicated behavior. Dependence of $N_{\rm eff}$ on $kT$ becomes steeper than linear especially for larger $kT$ (approximately $kT^{1.4}$) as higher $\ell$ start to contribute. Similarly, dependence on $D_0$ is only proportional to about $D_0^{-0.88}$, see Fig.~\ref{fig:effl_tot}. In general, different dependences with respect to those for individual $\ell$ is given by their contribution that changes for different $kT$ and $D_0$ (as indicated by Fig. \ref{fig:xs_ell1}). In reality, the dependence is more complicated than the simple power one. However, the deviation of actual $N_{\rm eff}$ from shown simple power dependence on $kT$ and $D_0$, as well as on the combination of $\Gamma_\gamma$ and $S_\ell$, is not huge and allows them to be effectively used. 

Note that although the ratio $N_{\rm eff}$ corresponding to neutrons with $\ell=0, 1,$ and $2$ is approximately 1:5:10, the corresponding actual resonance density has the ratios approximately 1:2.9:4.3. The difference in ratios comes from the fact that only a part of actual resonances for higher $\ell$ contributes to the MACS due to suppression of their contribution at low $E_n$ through the penetrability $P_\ell$. 
%We also verified that the $N_{\rm eff}$ for $\ell=0$ is close to the actual number of resonances $N$ in the energy interval of $(2-\sqrt{2}) kT < E_n < (2+\sqrt{2}) kT$, which correspond to two-sigma wide interval around the mean of the Maxwell-Boltzmann spectrum (mean $2kT$, standard deviation $\sqrt{2}kT$). 
%Deviation from exact validity is due to changes of the actual cross section with $E_n$.

%%%%%%%%%%%%%%%%%%%%%%%%%%%%%%%%%%%%%%%%%%%%%%%%%%%%%%%%%%%%%%%%%
\subsection{Fluctuation of MACS for individual $\ell$ \label{sec:art_part}}

\begin{figure}
\includegraphics[clip,width=\columnwidth]{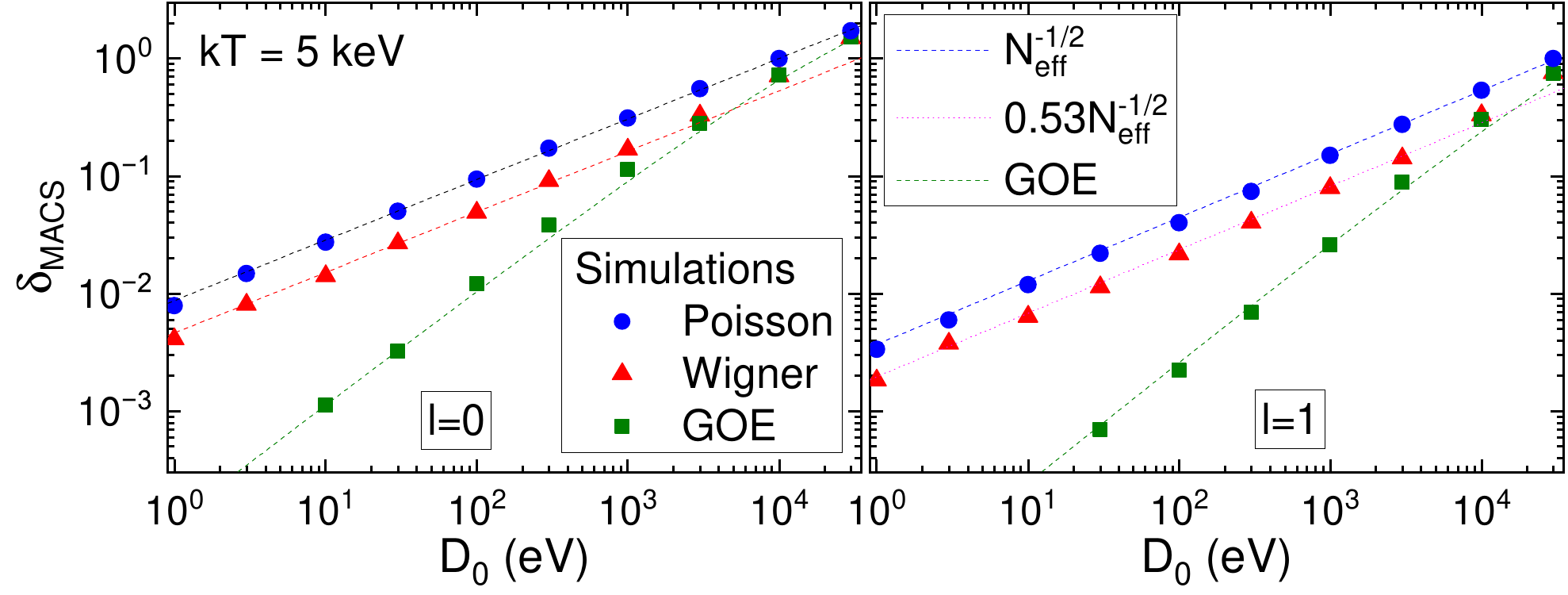}
\caption{Relative uncertainty in MACS, $\delta_{\rm MACS}$, due to different treatment of spacing for $\ell=0$ and 1 and $kT=5$~keV. $\Gamma_\gamma$ and $\Gamma_n$ do not fluctuate. 
}
\label{fig:spac1}
\end{figure}

\begin{figure*}
\includegraphics[clip,width=\textwidth]{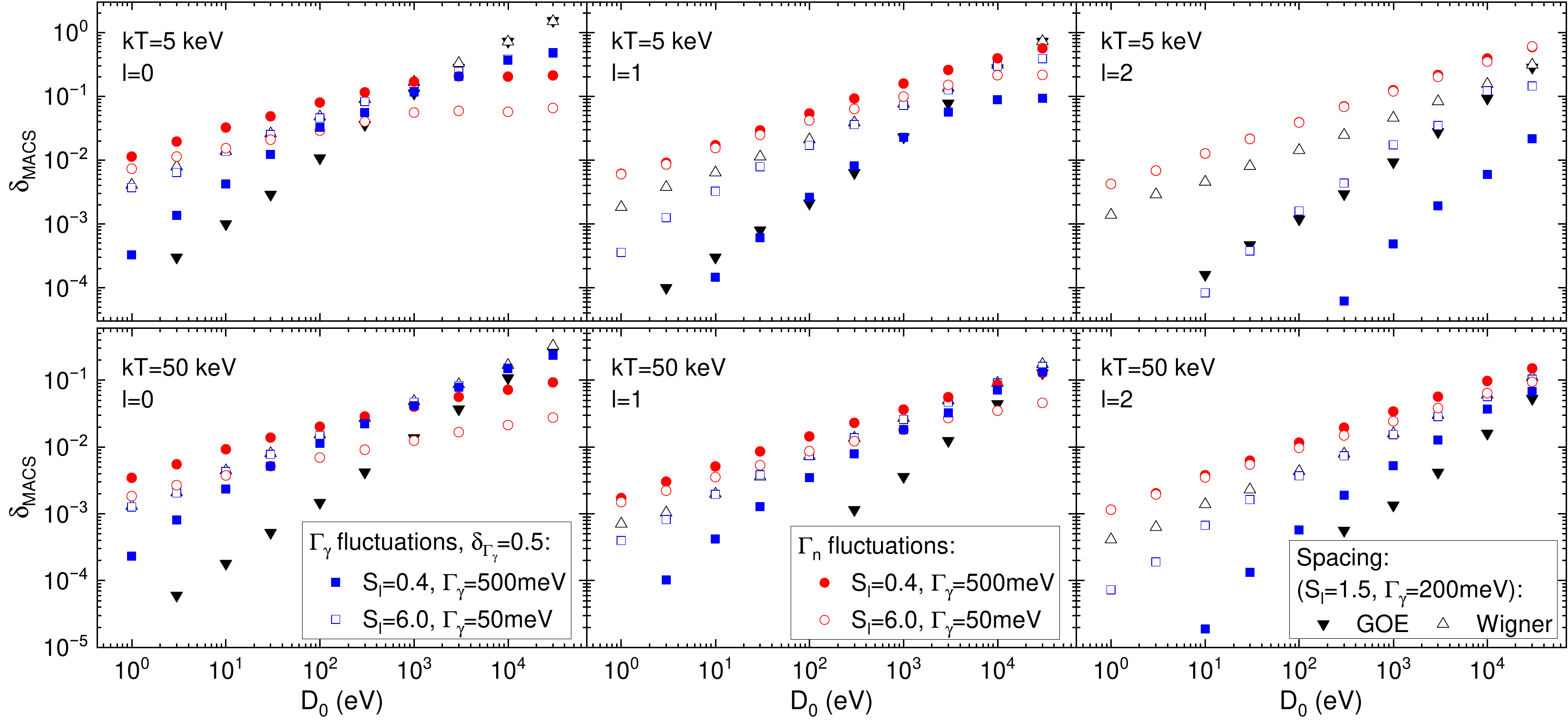}
\caption{Relative uncertainty in MACS, $\delta_{\rm MACS}$, due to different sources (spacing distribution, fluctuations in $\Gamma_n$ and $\Gamma_\gamma$) for artificial nucleus for $kT=5$ keV (top) and 50 keV (bottom) for individual $\ell=0-2$. All quantities but indicated one were fixed in simulations. 
}
\label{fig:test_var1}
\end{figure*}

The effective number of levels $N_{\rm eff}$ determined from the Poissonian NNS distribution above seems to be the relevant quantity for determination of fluctuation of MACS. Specifically, the relative uncertainty of MACS, $\delta_{\rm MACS}$, as a function of $D_0$ obtained with Wigner NNS distribution is (if $N_{\rm eff}$ is at least a few) fully consistent with $0.53 \times$ the value from Poissonian distribution that corresponds to the standard deviation of Wigner distribution. Similarly, $\delta_{\rm MACS}$ obtained from GOE sequences are nicely consistent with the expected fluctuation of number of resonances in an interval of selected length as derived from the number variance statistics \cite{Dyson63}, see Figs. \ref{fig:spac1} and \ref{fig:test_var1}. Once $N_{\rm eff}$ becomes very small, the fluctuations obtained using Wigner NNS start to deviate from the simple dependence ($0.53 \times$ Poisson) and become closer to the Poisson one. In practice, the value obtained with Wigner NNS nicely follows the GOE prediction here.

Further, if we switch off the fluctuation in resonance positions and check the fluctuations of MACS due to $\Gamma_\gamma$ or $\Gamma_n$, $\delta_{\rm MACS}$ in the limits $a$ or $b$ (where one of these quantities dominate the cross section) are very well described by expectations based on above-given values of $N_{\rm eff}$. 

Contributions from different sources of fluctuations to the relative uncertainty in MACS are shown at two different $kT$ in Fig. \ref{fig:test_var1}. The two ``extreme'' combinations of $S_\ell$ and $\Gamma_\gamma$ were used. The real nuclei should be thus confined between these two cases. The impact of fluctuation due to resonance positions is not shown for Poisson distribution of NNS. As indicated by Fig. \ref{fig:spac1}, with the exception of the largest $D_0$ where all NNS approaches merge, the Poisson NNS would give relative uncertainty about $2 \times$ larger than the value corresponding to Wigner NNS. The presented uncertainty in Fig. \ref{fig:test_var1} due to $\Gamma_\gamma$ fluctuations corresponds to rather large relative fluctuation of $\delta_{\Gamma_\gamma}=50\%$. This contribution scales well linearly with the value of $\delta_{\Gamma_\gamma}$ and it is thus smaller (compared to Fig. \ref{fig:test_var1}) in majority of nuclei, see Sec. \ref{sec:Ggestimate}. 

Figure \ref{fig:test_var1} indicates that various quantities may dominate $\delta_{\rm MACS}$ in different regions given by $D_0$, $S_\ell$, $\Gamma_\gamma$, and $kT$. In large areas of the space given by these quantities, the highest contribution  comes from fluctuation of $\Gamma_n$. However, there exist regions that are dominated by other quantities. For instance, for $\ell=0$ and $D_0\gtrsim 1000$ eV the most important contribution comes from fluctuations due to resonance spacing (with a significant contribution of fluctuation from $\Gamma_\gamma$).

%%%%%%%%%%%%%%%%%%%%%%%%%%%%%%%%%%%%%%%%%%%%%%%%%%%%%%%%%%%%%%%%%
\subsection{Effective neutron energy windows}

As a by-product of our analysis we checked expressions for effective neutron energy windows contributing to the MACS provided in the past by Wagoner~\cite{Wagoner69} and Rauscher {\it et al.}~\cite{Rauscher97}. The expressions given in Ref.~\cite{Rauscher97} for the center (mean energy) of the window $E_0$ and window width $\Delta$ (for $E_0$, $\Delta$, and $kT$ in keV) are:
\begin{eqnarray}
E_0 & \approx & 2 \times kT \times (\ell +1/2)
\label{eq:Rauscher_E} \\
\Delta & \approx & 2.26 \times kT \times (\ell +1/2)^{1/2}.
\label{eq:Rauscher_D}
\end{eqnarray}
Although Rauscher {\it et al.}~\cite{Rauscher97} cites Ref.~\cite{Wagoner69} as the source for these expressions, the multiplicative constants differ by a factor of two between the two references, with the Rauscher versions above being larger. Using our simulated sequences for $\ell \leq 3$ we arrived at results that slightly differ. Specifically, a very good empirical description of simulations was reached with:
\begin{eqnarray}
E_0 & \approx & 1.0 \times kT \times (\ell + 1.15)
%E_0 & \approx & 1.0 \times kT \times (\ell + 1.1)
\label{eq:MK_E} \\
\Delta & \approx & 0.68 \times kT \times (\ell + 2.8),
%\Delta & \approx & 0.5 \times kT \times (\ell + 4.4),
\label{eq:MK_D}
\end{eqnarray}
where $\Delta$ corresponds to two times the standard deviation of the distribution. These functions are only approximations as the actual results of the fit slightly depended on exact values of individual average resonance parameters adopted in simulations ($S_\ell$, $\Gamma_\gamma$, $D_0$). Figures comparing the simulated data with these relations
%as well as from predictions of Rauscher \cite{Rauscher97} and Wagoner \cite{Wagoner69} 
can be found in the Supplemental Material~\cite{SupplMat} together with an illustration of a shape of the contributions corresponding to all $\ell \leq 3$. Our values of $E_0$ and $\Delta$ are usually closer to those of Ref. \cite{Rauscher97} than Ref. \cite{Wagoner69}, but the difference could reach several tens of percent for different $\ell$.  

Note that the our dependence of $\Delta$ on $\ell$ differs from the previously suggested one.  Values of $\Delta$ from simulations do not show any sign of a square root dependence on $\ell$, particularly at higher temperatures. Instead, a linear dependence on $\ell$ seems to be a very good approximation. This is illustrated in the Supplemental Material. Comparison of the effective number of levels $N_{\rm eff}$, given by Eq. (\ref{eq:Neff}), with the actual number of levels $N_\Delta$ within interval $\Delta$ indicates no simple relation between these two quantities as a function of $\ell$. In practice, the ratio $N_{\rm eff}/N_\Delta$ varies approximately between 0.9 and 2.3 for $D_0=1-10000$~eV and $\ell\leq 2$. 
%Even larger spread of the ratio is reached when the actual number of resonances $N$ is taken from two-sigma wide interval around the mean of the Maxwell-Boltzmann spectrum (mean $2kT$, standard deviation $\sqrt{2}kT$).

%%%%%%%%%%%%%%%%%%%%%%%%%%%%%%%%%%%%%%%%%%%%%%%%%%%%%%%%%%%%%%%%%
\subsection{Total fluctuation of MACS from major sources \label{sec:art_tot}}

\begin{figure}
\includegraphics[clip,width=\columnwidth]{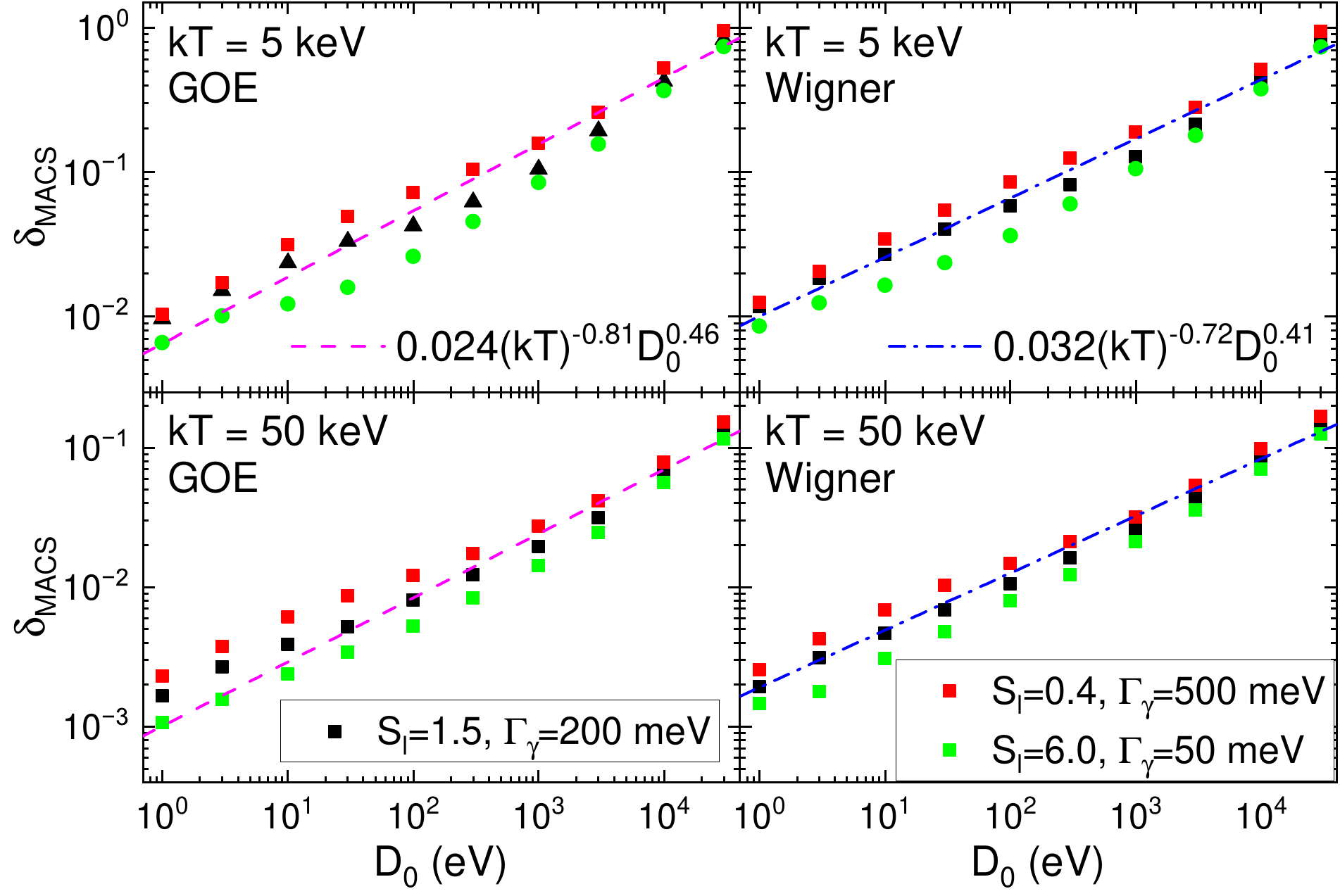}
\caption{Relative uncertainty in MACS, $\delta_{\rm MACS}$, for the artificial nucleus considering all $\ell$. Top panels correspond to $kT=5$ keV, while bottom ones to $kT=50$ keV; left panels to resonance positions from GOE sequences, while the right ones to Wigner NNS. The lines correspond to dependence obtained from fits to real nuclei, see Eq.~(\ref{eq:delta_final}) in Sec.~\ref{sec:actual_macs}. Fluctuation of $\Gamma_\gamma$ corresponds to the dependence proposed in Sec.~\ref{sec:Ggestimate}.}
\label{fig:unc_art_tot1}
\end{figure}

Although the contributions to MACS fluctuations discussed above are not complete, they are important and it is worthwhile to check the sum of these contributions for all $\ell$, i.e. their total contribution to $\delta_{\rm MACS}$. It is not guaranteed that a simple dependence on $D_0$ and $kT$ can be found because (as indicated above, see, e.g., Fig.~\ref{fig:xs_ell1}) the contribution of different $\ell$ can significantly vary. Figure~\ref{fig:unc_art_tot1} shows this summed contribution for two different $kT$ and three combinations of $S_\ell$ and $\langle \Gamma_\gamma \rangle$. The relative fluctuation in $\Gamma_\gamma$, $\delta_{\Gamma_\gamma}$, used in the figure corresponded to the dependence suggested in Sec.~\ref{sec:Ggestimate}.  

%Despite some impact of $S_\ell$ and $\Gamma_\gamma$ on fluctuations of MACS, a simple general dependence of $\delta_{\rm MACS}$ on $D_0$ can be seen in Figure~\ref{fig:unc_art_tot1}. At the same time the figure indicates that the true relative uncertainty can not be exactly described with a simple power function of $D_0$. The relative uncertainty of MACS for a specific nucleus thus can probably be estimated only within a factor of about two, if knowledge of $S_\ell$ and $\Gamma_\gamma$ is missing and we know only $D_0$. 
While  $S_\ell$ and $\Gamma_\gamma$ combine to impact fluctuations in the MACS, a simple general dependence of $\delta_{\rm MACS}$ on $D_0$ can be observed in Figure~\ref{fig:unc_art_tot1}. At the same time, the figure indicates that the true relative uncertainty can not be described solely as a power function of $D_0$. The relative uncertainty in MACS for a specific nucleus could thus be estimated within a factor of about two if only $D_0$ is know.  A more precise estimate clearly requires knowledge of $S_\ell$ and $\Gamma_\gamma$. 

%%%%%%%%%%%%%%%%%%%%%%%%%%%%%%%%%%%%%%%%%%%%%%%%%%%%%%%%%%%%%%%%%
\subsection{Impact of additional effects on fluctuation of MACS}

Fluctuation properties of MACS can be influenced also by other sources discussed in detail below. They include possible contribution of other outgoing channels and several effects influencing density of resonances: spin of the target nucleus $J_T$, spin dependence of resonance density (often described by a spin cut-off parameter $\sigma_c$) as well as excitation-energy dependence of resonance density. The fluctuations from all these individual sources should be independent and thus summed in squares. 

%In practice, to test MACS fluctuations due to these effects we mainly made simulations with resonances fixed at positions given by average spacing and assuming no fluctuation of individual widths. For some tests we also kept the sequence of neutron widths (generated from Porter-Thomas distribution) the same for all tested resonance sequences.

%%%%%%%%%%%%%%%%%%%%%%%%%%%%%%%%%%%%%%%%%%%%%%%%%%%%%%%%%%%%%%%%%
\subsubsection{Impact of resonance density \label{sec:art_dens}}

The impact of the spin dependence of resonance spacing was checked using different spin cut-off parameter $\sigma_c$. Specifically, three different expressions (BSFG and CT models from \cite{Egidy05,Egidy09}) were adopted (and $J_T=0$ assumed). 

In general, as we exploit the dependence as a function of $D_0$ one might expect that different expressions for $\sigma_c$ can play larger role for spins farther from the capturing spin, i.e. larger $\ell$. Nonetheless, only a small impact on $\delta_{\rm MACS}$ was observed with these expressions; the default $\sigma_c$ gave the smallest $\delta_{\rm MACS}$. The relative change (increase) in $\delta_{\rm MACS}$ due to different $\sigma_c$ is at most a few per cent for $\ell<2$ and up to about 10\% for $\ell=2$.  

Effects of a similar size were observed if we checked impact of different $T_{CT}$ that characterizes the derivative of the resonance density. The value of $T_{CT}$ for actual tested nuclei, see Sec. \ref{sec:actual}, spans the range between $0.56$ MeV$^{-1}$ ($^{156}$Gd) and $1.2$ MeV$^{-1}$ ($^{63}$Ni) \cite{Egidy05}. Change of $T_{CT}$ within this range has virtually no impact for $kT \lesssim 40$~keV as the change of resonance density is similar for all $T_{CT}$ values, while for $kT=90$ keV the difference in $\delta_{\rm MACS}$ between the minimum or maximum tested $T_{CT}$ with respect to the value used for the artificial nucleus ($T_{CT}=0.7$ MeV$^{-1}$) is $\approx 10\%$. Larger value of $T_{CT}$ results in larger $\delta_{\rm MACS}$. 

We further performed tests with ``BSFG-like'' energy dependence of the density of resonances, adopting the spin cut-off parameter for BSFG model of Ref.~\cite{Egidy05}. In general, the BSFG-like dependence yields smaller increase in resonance density with $E_n$, thus slightly increases $\delta_{\rm MACS}$. The impact of this energy dependence is again small for $kT \lesssim 40$ keV, while it starts to be visible for larger $kT$; $\delta_{\rm MACS}$ is larger by about 15\% for $kT=90$ keV. Note that all three effects discussed so far in this section have similar impact independently of  checked $D_0$ values.

Finally, for nuclei with $J_T>0$ there are additional effects that come into play. First, the contribution of resonances with different spins to the cross section might become slightly different as a result of the interplay between the statistical factor $g_J$ and the spin dependence of resonance spacing. 
%This effect could slightly increase $\delta_{\rm MACS}$. 
Second, the fluctuation of $\Gamma_n$ for a fraction of resonances -- namely those with "middle" allowed spin for neutrons with $\ell>0$ -- are not governed by the $\chi^2_{\nu=1}$ (Porter-Thomas) but by $\chi^2_{\nu=2}$ (exponential) distribution. 
Third, the NNS distributions discussed above correspond to resonance sequences of the same $J^\pi$. A mixture of more sequences slightly changes the actual distributions -- there are two $J^\pi$ for $\ell=0$ and more $J^\pi$ than for $J_T=0$ for $\ell>0$. We tested the impact of these effects for several $J_T$ up to $9/2$.
All these three effects associated with $J_T>0$ lead to increase of $\delta_{\rm MACS}$. This increase for specific $\ell$ can be up to a factor of about 1.4. 
%(the impact of $\ell=2$ resonances might be even higher especially for $D_0\gtrsim 1000$ eV ... OK). 
Nonetheless, the combined effect from all $\ell$ was found to be smaller than a factor of $\approx 1.25$. 

Combination of all the effects related to spacing of resonances can then lead to an increase of $\delta_{\rm MACS}$ with respect to the results obtained from previous subsections at maximum by a factor of $\approx 1.5$. If the target nucleus has $J_T=0$, the possible increase in $\delta_{\rm MACS}$ is about a half of this value. These effects thus should not be large enough to prevent providing an effective practical estimation of $\delta_{\rm MACS}$ of a simple form.

%%%%%%%%%%%%%%%%%%%%%%%%%%%%%%%%%%%%%%%%%%%%%%%%%%%%%%%%%%%%%%%%%
\subsubsection{Additional decay channels \label{sec:art_inel}}

Thus far we have limited ourselves to consideration of cases where elastic scattering and radiative capture are the only open decay channels.  Depending on the nucleus, additional decay modes of a resonance might be open, particularly as incoming neutron energy increases. The channels most likely to be open include the (n,$\alpha$) reaction in light nuclei, fission (in actinides), or inelastic neutron scattering. Practically, light nuclei often cannot be described by a statistical approach. Further, relevance of MACS is restricted in actinides as they are formed only in $r$-process. Given these conditions, we will consider only inelastic scattering as it is often the most relevant additional channel. Nonetheless, the general conclusions are similar for other additional reaction channels.  Inelastic scattering can start to contribute once the neutron energy exceeds the first excited state of the target nucleus, which can be as low as a few keV. 

Once inelastic scattering is included, the resonance kernel has the form
\begin{equation}
R_k = \frac{\Gamma_n \Gamma_\gamma}{\Gamma_n + \Gamma_\gamma + \sum\limits_{n'}\Gamma_{n'}},
\end{equation}
where $\Gamma_{n'}$ indicates inelastic neutron scattering populating a single level. In principle, the last term in denominator could correspond to sum of all other reaction channels. 

If the possible coupled-channel effect is not considered, $\Gamma_{n'}$ should fluctuate (similarly to $\Gamma_n$) around its expectation value according to $\chi^2_\nu$ distribution with $\nu=1$ or 2, depending on spins and parities of the resonance and the final level. If the number of levels contributing to the inelastic scattering was very large, the distribution of $\sum\Gamma_{n'}$ would become a Normal distribution. The fluctuation properties of sum of all additional channels should lie between these two limits. 

The uncertainty of MACS due to $\sum\Gamma_{n'}$ then depends on ratios of expectation values of all quantities involved in $R_k$ as well as the exact fluctuation properties of $\Gamma_\gamma$ and $\sum\Gamma_{n'}$. In all cases, the presence of inelastic scattering increases relative fluctuations of $R_k$, sometimes significantly. 
On the other hand, the absolute value of $R_k$ (and thus the contribution of the cross section at given neutron energy to MACS) is lowered by inelastic scattering.

We tested the behavior of fluctuations of $R_k$ assuming that $\sum\Gamma_{n'}$ comes from two distributions, including $\chi^2_{\nu=1}$ and a Normal one with a few $\delta_{\Gamma_{n'}}$ ranging from 0.05-0.4, for a broad range of realistic expectation values of involved decay widths. It seems difficult to get any simple quantification of the expected fluctuations. We only found that the quantity $\langle R_k \rangle \times \sigma_{R_k}$, where $\sigma_{R_k}$ is the standard deviation of the capture kernel distribution, remains approximately constant (could change at most by a factor of a few) for a broad range of realistic ratios $\sum_{n'}\langle \Gamma_{n'} \rangle / \langle \Gamma_{n} \rangle \lesssim 10^{-1}-10^{0}$. For $\sum_{n'}\langle \Gamma_{n'} \rangle \gg \langle \Gamma_{n} \rangle$ the quantity is then much smaller than its value in the absence of inelastic scattering. 

These findings mean that the {\em absolute} value of the MACS uncertainty when an inelastic channel is open should not be significantly higher than the uncertainty of MACS without considering any inelastic channel. However, any simple quantification of actual impact on $\delta_{\rm MACS}$ is difficult due to the suppression of $R_k$ and resulting decrease of the MACS expectation value. In practice, an impact of the inelastic scattering on $\delta_{\rm MACS}$ can be checked only on real nuclei as it strongly depends on excitation energies and $J^\pi$ of involved states.

%%%%%%%%%%%%%%%%%%%%%%%%%%%%%%%%%%%%%%%%%%%%%%%%%%%%%%%%%%%%%%%%%
\subsection{Comparison with real nuclei \label{sec:actual}}

\begin{table*}
\begin{threeparttable}
\begin{tabular}{ccccccccc}
\hline \hline
Quantity                                                          & $^{63}$Ni & $^{65}$Cu& $^{77}$Se& $^{87}$Sr& $^{96}$Zr & $^{156}$Gd& $^{157}$Gd& $^{169}$Tm\\
\hline
\multicolumn{9}{c}{Adopted values} \\
\hline
$J^\pi_T$                                                         &1/2$^-$    & 3/2$^-$  & 1/2$^-$  & 9/2$^+$  &  0$^+$    & 0$^+$     & 3/2$^-$   & 1/2$^+$  \\ 
$D_0$ (eV)                                                        & 4400      & 1500     & 110      &  290     &  13000    &  30       & 4.9       & 7.3      \\
$S_0$ ($\times 10^{-4}$)                                          & 2.0       & 2.16     & 1.49     &  0.44    &  0.34     &  2.0      & 2.2       & 1.72     \\
$S_1$ ($\times 10^{-4}$)                                          & 0.5       & 0.38     & 4.3      &  2.7     &  6.0      &  1.5      & 1.5       & 1.09     \\
$S_2$ ($\times 10^{-4}$)                                          & 1.0       & 1.0      & 2.0      &  2.0     &  1.1      &  3.0      & 3.0       & 3.0      \\
$\langle\Gamma_\gamma^{\ell=0}\rangle$ (meV)                      & 254/536   & 387/403  & 350/416  &  155/147 &  71       &  88       & 104/99    &  97/96   \\
$\langle\Gamma_\gamma^{\ell=1}\rangle$                            & 424-703   & 309-360  & 358-384  & 149-174  &  64-87    &  87-88    & 97-104    &  93-95   \\
$T_{CT}$ (MeV$^{-1}$)~\cite{Egidy05}                                   & 1.2       & 1.12     & 0.93     &  0.96    &  0.7      & 0.56      & 0.58      & 0.61     \\
\hline
\multicolumn{9}{c}{Literature values from Refs.~\cite{Mughabghab18a,Mughabghab18b}} \\
\hline
$D_0$ (eV)                                    & 4400(800) & 1520(100)& 110(11)  &  268(54) &12800(2000)& 30.5(1.7) & 4.65(30)  & 7.28(43) \\
$S_0$ ($\times 10^{-4}$)                      & -         & 2.16(37) & 1.49(38) & 0.50(18) & 0.34(14)  & 1.95(36)  & 2.2(2)    & 1.72(12) \\
$S_1$ ($\times 10^{-4}$)                      & -         & 0.38(5)  & 4.3(10)  & 2.63(57) & 6.0(18)   & 1.7(3)    & 2.2(7)    & 1.09(7)  \\
$S_2$ ($\times 10^{-4}$)                      & -         & -        & -        & -        & -         & 2.6(4)    & -         & 3.05(30) \\
$\langle\Gamma_\gamma^{\ell=0}\rangle$ (meV)  &1390(8)\tnote{u}& 395(40)  & 390(54)  &  150(30) &  71(8)    &  79(1)    & 105(3)    &  120(7)  \\
$\langle\Gamma_\gamma^{\ell=1}\rangle$ (meV)  & -         & 260(15)  & 418(26)  &170(24)\tnote{u}& 151(75)  &  44(1)\tnote{u}& 57(4)\tnote{u} & 65(6)\tnote{u}\\
\hline
\multicolumn{9}{c}{Literature values from Ref.~\cite{RIPL3}} \\
\hline

$D_0$ (eV)                                    & -         & 1300(110)& 120(15)  &  290(80) &13000(3000)&  30(6)    & 4.9(5)    & 8.5(7)   \\
$S_0$ ($\times 10^{-4}$)                      & -         & 2.20(30) & 1.20(20) & 0.34(6)  & 0.30(15)  & 1.90(40)  & 2.20(40)  & 1.50(20) \\
$S_1$ ($\times 10^{-4}$)                      & -         & 0.47(8)  & 4.3(10)  & 2.7(6)   & 9.0(20)   & -         & -         & -        \\
$\langle\Gamma_\gamma^{\ell=0}\rangle$ (meV)  & -         & 385(20)  & 390(60)  &  150(40) &  65(15)   &  88(12)   & 97(10)    &  97(10)  \\
$\langle\Gamma_\gamma^{\ell=1}\rangle$ (meV)  & -         & 260(80)  & -        &  -       &  120(75)  &  -        & -         &  -       \\
\hline
MACS $@$ 30 keV, $\ell=0/1/2$ $(\%)$\tnote{a} \hspace{3mm} & 33/57/10  & 52/35/12 & 38/58/4  & 33/55/12 & 12/54/30  & 42/52/6   & 61/35/4   & 55/39/5  \\
\hline\hline
\end{tabular}
\begin{tablenotes}
\item[a] The fraction of MACS for $kT=30$ keV corresponding to neutrons with different (indicated) orbital momentum.  
\item[u] Not derived from individual resonance parameters but from the cross section, see Refs. \cite{Mughabghab18a,Mughabghab18b}.  
\end{tablenotes}
\end{threeparttable}
\caption{\label{tab:parameters}
Values of average resonance parameters used in simulations of real nuclei (top part). They are dominantly based on values from Mughabghab~\cite{Mughabghab18a,Mughabghab18b} or  RIPL-3~\cite{RIPL3} that are also listed (middle parts); adopted $S_2$ values are also based on Figure 2.3 of Refs. \cite{Mughabghab18a,Mughabghab18b}. The fraction of $kT=30$ keV MACS corresponding to neutrons with different orbital momentum is also listed. The two values of $\langle\Gamma_\gamma^{\ell=0}\rangle$ in "Adopted values" (top) for odd-mass nuclei ($J_T\neq 0$) correspond to the values from the two spins of $s$-wave resonances. The values are means of the distributions obtained from DICEBOX simulations; a common renormalization factor was applied to achieve average simulated values near experimental ones. The range for $\langle\Gamma_\gamma^{\ell=1}\rangle$ corresponds to minimum and maximum simulated averages for individual resonance spins; the same renormalization factor as for $\langle\Gamma_\gamma^{\ell=0}\rangle$ was applied. 
}
\end{table*}

So far we have studied $\delta_{\rm MACS}$ in an artificial nucleus as a function of $D_0$ and often for individual $\ell$. To check the relevance of above-given results for nature we need to consider real nuclei, each of which has a particular, definite value of $D_0$ and the MACS is given by unique combination of contributions from individual $\ell$. 
To reach this goal we simulated the MACS for eight different target nuclei, namely $^{63}$Ni, $^{65}$Cu, $^{77}$Se, $^{87}$Sr, $^{96}$Zr, $^{156}$Gd and $^{157}$Gd, and $^{169}$Tm. They cover $D_0$ range spanning more than three orders of magnitude and also a variety of $J_T$, $S_l$, and $\langle \Gamma_\gamma \rangle$. This set includes nuclei with both even and odd $Z${} and ${N}$ (no odd-$Z$, odd-$N$ nuclei were included due to very restricted number of them with available data). A broad range of $D_0$ allows comparison of the dependence on $D_0$ obtained for artificial nucleus. Basic characteristics of these nuclei are listed in Tab. \ref{tab:parameters}. A few of these nuclei show contribution of inelastic scattering from rather low energies. This is especially the case of $^{169}$Tm where the inelastic channel opens already at about 8~keV.

%%%%%%%%%%%%%%%%%%%%%%%%%%%%%%%%%%%%%%%%%%%%%%%%%%%%%%%%%
\subsubsection{Estimate of $\Gamma_\gamma$ fluctuations  \label{sec:Ggestimate}}

\begin{figure}
\includegraphics[clip,width=0.95\columnwidth]{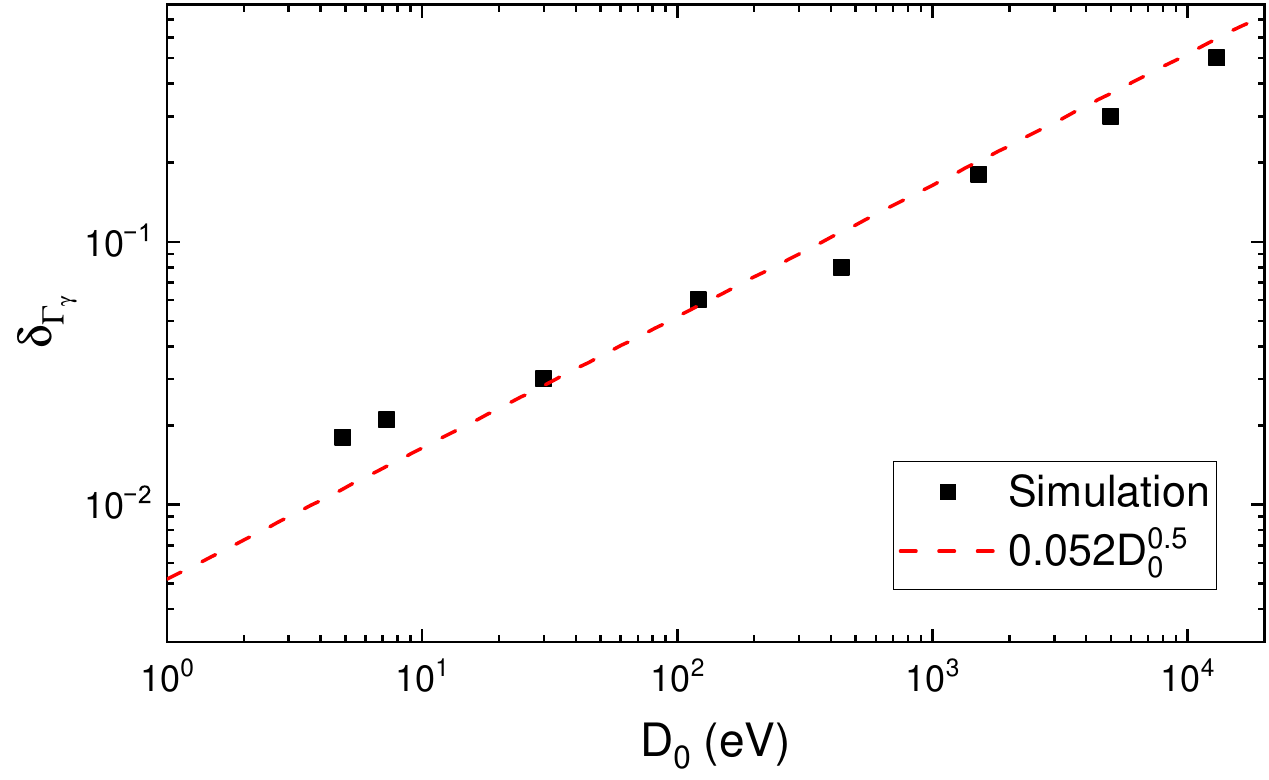}
\caption{Relative uncertainty for simulated distribution of $\Gamma_\gamma^{\ell=0}$, $\delta_{\Gamma_\gamma}$ for real nuclei plotted as a function of $D_0$. Individual values correspond to an average of distribution for all $s$-wave resonances as simulated by DICEBOX. 
}
\label{fig:Gg_rel_unc}
\end{figure}

Fluctuation of $\Gamma_\gamma$ should to high degree depend on level density of the nucleus and thus resonance spacing. To check this dependence we exploited $\Gamma_\gamma$ distribution simulated with DICEBOX \cite{Becvar98} code for all isotopes listed in Tab. \ref{tab:parameters}. The relative uncertainty $\delta_{\Gamma_\gamma}$, estimated from $\ell=0$ resonances, for each checked nucleus is shown in Fig. \ref{fig:Gg_rel_unc} as a function of $D_0$. For nuclei with larger $D_0$, $\delta_{\Gamma_\gamma}$ can depend on the resonance $J^\pi$ and the differences in individual $\delta_{\Gamma_\gamma}$ can reach about 20-30\%. 
We would like to stress here that detailed description of the simulation of $\Gamma_\gamma$ and discussion of $\Gamma_\gamma$ fluctuation properties would be rather lengthy and goes beyond the scope of this paper. For determination of $\delta_{\rm MACS}$ it is sufficient if we find an approximate dependence of $\delta_{\Gamma_\gamma}$ on $D_0$. Figure \ref{fig:Gg_rel_unc} indicates that although details of such dependence might be more complex, there is a clear general that can be approximated with 
\begin{equation}
\delta_{\Gamma_\gamma} \equiv \sigma_{\Gamma_\gamma}/ \langle \Gamma_\gamma \rangle \approx 5.2 \times 10^{-3} (D_0)^{1/2}
\end{equation}
where $D_0${} is in eV.
%{\it Or we can say that $D_0$ is in eV here (and remove ``1 eV'').} 
Although this dependence is not an exact description, it gives a reasonable and useful estimation of $\delta_{\rm \Gamma_\gamma}$.

A proportionality of $\delta_{\Gamma_\gamma}$ to $\approx D_0^{1/2}$ should not be very surprising as the resonance spacing reflects the level density also below $S_n$. Resonance spacing thus  should be (at least approximately) inversely proportional to the number of $\gamma$-ray transitions by which the resonance can decay and $\delta_{\Gamma_\gamma}$ can then be expected to be approximately inversely proportional to the square root of the number of these transitions. 

%%%%%%%%%%%%%%%%%%%%%%%%%%%%%%%%%%%%%%%%%%%%%%%%%%%%%%%%%%%%%%%%%
\subsubsection{Fluctuation of MACS \label{sec:actual_macs}}

\begin{figure}
\includegraphics[clip,width=\columnwidth]{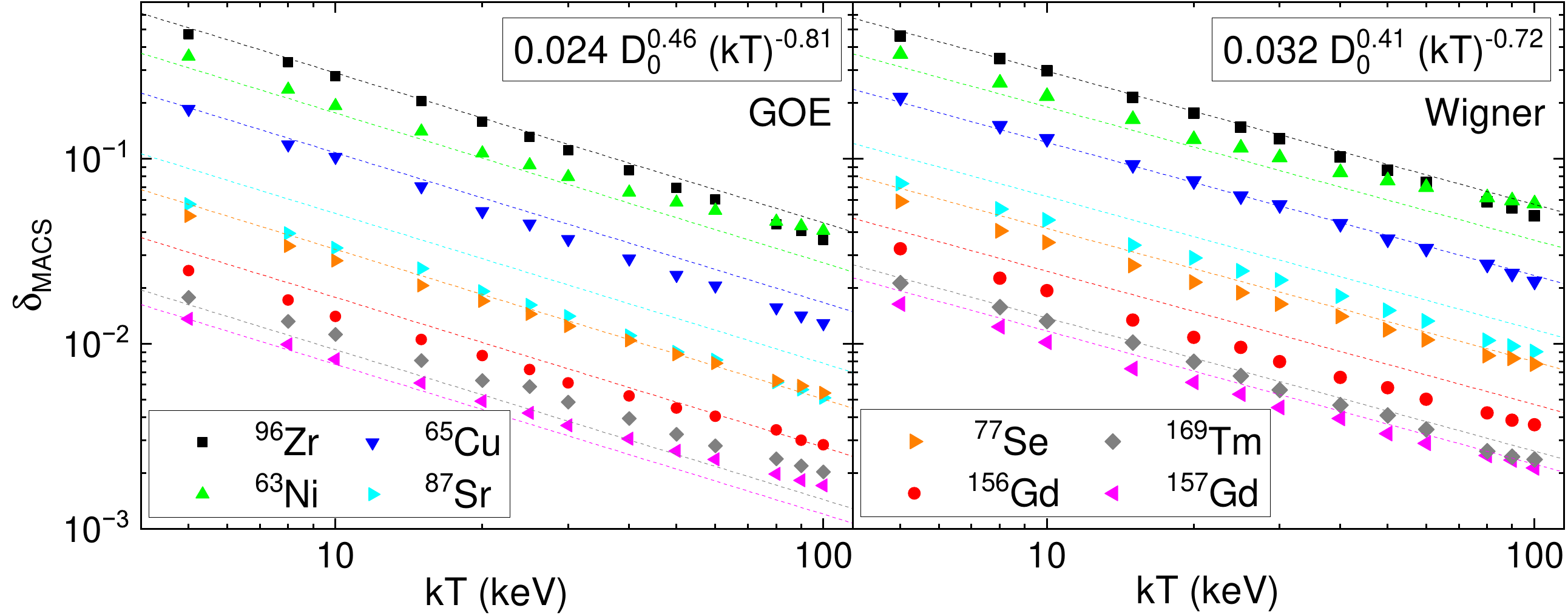}
\caption{Relative uncertainty $\delta_{\rm MACS}$ for real nuclei using resonance positions generated from GOE sequences (left) and Wigner NNS (right).}
\label{fig:xs_actual}
\end{figure}

%The expected relative fluctuations of MACS for artificial nuclei was already discussed in Sec. \ref{sec:art_tot}. 
Let us now check whether the results for the artificial nucleus derived above are consistent with the behavior of real nuclei where we have a contribution from all effects discussed in previous subsections.  
The relative uncertainty $\delta_{\rm MACS}$ for all eight nuclei as a function of $kT$ is shown in Fig.~\ref{fig:xs_actual} for NNS governed by Wigner distribution and GOE-based sequences. Individual $\Gamma_\gamma$ fluctuated according to the actual distribution from DICEBOX simulations. The lines correspond to the global fit of the form $\delta_{\rm MACS} = C_1 \times D_0^{C_2} \times (kT)^{C_3}$.
Fit parameters for the different treatment of resonance positions are:
\begin{eqnarray}
\delta_{\rm MACS} & = & 0.032 D_0^{0.41} (kT)^{-0.72} \quad {\rm for\ Wigner\  NNS} 
\label{eq:delta_final}\\
\delta_{\rm MACS} & = & 0.024 D_0^{0.46} (kT)^{-0.81} \quad {\rm for\ GOE\ sequences} \nonumber 
%\\
%\delta_{\rm MACS} & = & 0.032(3) D_0^{0.41(2)} (kT)^{-0.72(2)} \label{eq:delta_final}\\
%\delta_{\rm MACS} & = & 0.024(3) D_0^{0.46(2)} (kT)^{-0.81(2)} \nonumber
\end{eqnarray}
with $D_0$ in eV, while $kT$ in keV. 

Although the simple functions used (that depend only on $D_0$ in addition to $kT$) do not reproduce individual points perfectly,
they seem to be sufficient for all practical purposes. The maximum deviation of points is by a factor of about 1.5, i.e. by the value which is even smaller than that anticipated in Sec. \ref{sec:art_tot}. 
 
Deviations from the simple fit dependence start to appear mainly (although not exclusively) for higher $kT$ where inelastic scattering can play a significant role. Nonetheless, the impact of this effect is limited and an approximation via a simple expressions of Eq. (\ref{eq:delta_final}) is reasonably valid even if the inelastic channels are opened.
%In line with the results presented in Sec. \ref{sec:art_inel}, the relative uncertainty $\delta_{\rm MACS}$ is slightly larger here, but still within the above-quoted range. 
 
The use of a BSFG-like energy dependence of resonance density yields results similar to the CT-like one. The impact of BSFG was discussed more in Sec. \ref{sec:art_dens}, and it effectively gives a slightly weaker dependence of $\delta_{\rm MACS}$ on $kT$, but the impact is very small (well within uncertainties of fit), being negligible for $kT \lesssim 40$ keV. 
%{\it Fits are (I have not run simulations for GOE, but I can): Poi: all: $0.0468(50) D_0^{0.392(11)} (kT)^{-0.644(17)}$, below 30 keV: $0.0464(68) D_0^{0.392(15)} (kT)^{-0.642(30)}$; Wig: all: $0.0288(33) D_0^{0.422(12)} (kT)^{-0.698(18)}$, below 30 keV: $0.0294(46) D_0^{0.422(16)} (kT)^{-0.707(31)}$.}

\begin{figure*}
\includegraphics[clip,width=\textwidth]{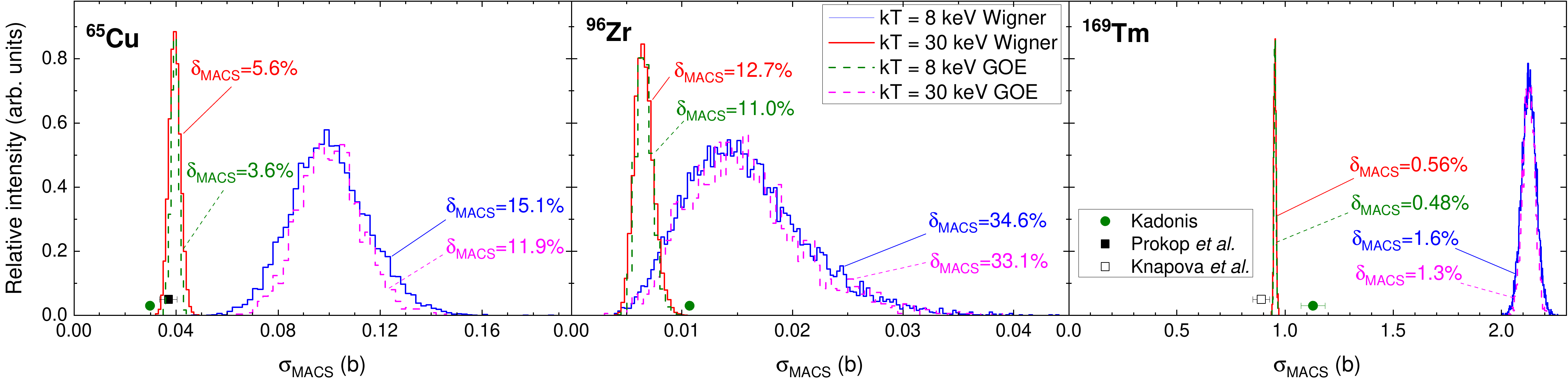}
\caption{(Color online) Distribution of simulated MACS for astrophysically relevant $kT=8$ and 30~keV for three checked nuclei. The values  of $\delta_{MACS}$ deduced from simulated distribution are provided together with experimental data for $kT=30$~keV. Experimental data are from Kadonis~\cite{Kadonis}, and from newer works of Prokop {\it et al.}~\cite{Prokop19} and Knapov\'a {\it et al.}~\cite{Knapova25} for Cu and Tm, respectively.
}
\label{fig:macs2}
\end{figure*}

The parametrization of Eq. (\ref{eq:delta_final}) is also shown in Fig.~\ref{fig:unc_art_tot1} indicating that $\delta_{\rm MACS}$ for real nuclei is well consistent with results obtained from tests with the artificial nucleus. Illustration of the expected distribution of MACS for $kT=8$ and 30 keV is shown for three isotopes in Fig. \ref{fig:macs2}.

Experimentally determined MACS for $kT=30$ keV are also shown in Fig.~\ref{fig:macs2} for comparison. The experimental values are reasonably, not perfectly, reproduced with the predicted distributions. Although our aim is not to discuss \emph{absolute} values of MACS (including agreement of simulated and experimental value) in detail in this paper, we comment briefly on the agreement of absolute MACS. In general, inconsistency of experimental data with simulated distribution comes from use of inappropriate average resonance parameters in the latter. As evident from Table~\ref{tab:parameters}, $\Gamma_\gamma^{\ell=1}$ values obtained from DICEBOX simulations for $^{65}$Cu and $^{96}$Zr do not fully reproduce the values available in evaluations \cite{Mughabghab18a,Mughabghab18b,RIPL3}. Experimental points for both isotopes would be nicely consistent with the predicted distribution if the used $\Gamma_\gamma^{\ell=1}$ were correspondingly changed (as a non-negligible part of the cross section comes from neutrons with $\ell=1$, see the fraction in Table \ref{tab:parameters}). Problems with $D_0$ and $\Gamma_\gamma$ from evaluations are also present for $^{169}$Tm. They are thoroughly discussed in Ref. \cite{Knapova25}. If we adopted the  $\langle\Gamma_\gamma^{\ell=0}\rangle=86$ meV proposed in ENDF-VII.0, or larger $D_0$ as suggested in \cite{RIPL3}, the predicted distribution would be perfectly consistent with the experimental value from \cite{Knapova25}. 

We make one additional note connected with the absolute value of MACS and the applicability of statistical approach. When the number of resonances contributing to the MACS becomes small, HF predictions of MACS are typically expected to be unreliable, see Ref.~\cite{Rochman13} for a couple of examples. Rauscher {\it et al.} \cite{Rauscher97} provide a rule-of-thumb, estimating that the HF deviation falls below 20\% if there are at least $5-10$ resonances within the effective energy window $\Delta$. If one considers $kT$ larger than a few keV, relevant for astrophysical processes, this number of resonances corresponds to $D_0$ high enough that the contribution of $\ell\ge1$ resonances to MACS is significant. About 10 resonances within $\Delta$ is then reached for $D_0\approx kT$ when the $d$-wave contribution is small. 
Using Eq.~(\ref{eq:delta_final}) one finds that $D_0=kT$ corresponds to $\delta_{\rm MACS}\approx 0.3$, 0.2, and 0.13 for $kT=8, 25$, and 90 keV.
%, while $D_0=5kT$ to $\delta_{\rm MACS}\approx 0.55$, 0.4, and 0.25, respectively. {\it Or should we list only 5kT values?} 
Such values of $\delta_{\rm MACS}$ mean that a use of HF prediction for determining MACS could be problematic under these conditions.  From this analysis, we observe that the 5-10 levels rule-of-thumb from \cite{Rauscher97} is somewhat optimistic; with ten levels within the energy window, we observed variations of 20\% at $kT=25$~keV, with larger variations at lower energies/temperatures.

If we consider even larger level spacings such as $D_0=5kT$, we find $\delta_{\rm MACS}\approx 0.55$, 0.4, and 0.25, for the same $kT=8, 25$, and 90~keV discussed above.
In such cases, the MACS distribution becomes highly asymmetric.
If $kT\ll D_0$, the vast majority of sequences yield a very low MACS but a few result in a MACS at comparable with, and very occasionally significantly larger than, HF prediction.  An illustration of the behavior of distribution of MACS from individual generated sequences for $kT\ll D_0$ can be found in the Supplemental Material~\cite{SupplMat}.

%%%%%%%%%%%%%%%%%%%%%%%%%%%%%%%%%%%%%%%%%%%%%%%%%%%%%%%%%%%%%%%%%
\subsection{Impact of measured cross section at low $E_n$}

Information on the cross section and resonance parameters is often only available below a certain measured neutron energy. This is a result of the fact that the flux and neutron-energy resolutions at TOF facilities typically decreases at increasing neutron energy. In order to understand how such a partial measurement would impact the corresponding uncertainty in the MACS, we considered a hybrid model where information on the cross section was precisely known below some maximum energy $E_m$ and only average resonance parameters were available above $E_m$. We then first checked relative uncertainty of MACS considering only the region of energies $E_n>E_m$. It was found that the relative uncertainty of this MACS fraction 
does not significantly change with respect to $\delta_{MACS}$. More precisely, it slightly decreases for larger $E_m$. However, this decrease is only by about 10\% for $E_m \approx kT$ and reaches at maximum about 25\% for maximum checked $E_m=50$ keV (slightly depending on $kT$ and the real nucleus). 
%{\it Do we have any clear explanation? This indicates that ``effective number of levels'' involved is almost constant/slightly increases with neutron energy. Is this what we should expect? I am not really sure -- I would probably expect that (if the cross section was more or less constant over the region and the level density was similar -- the case if we do not have very high $kT$) a reduction of considered interval should lead to reduction of effective number of levels and thus increase of relative uncertainty. The actual behavior is likely somehow connected with the shape of the neutron flux.} 

\begin{figure}
\includegraphics[clip,width=\columnwidth]{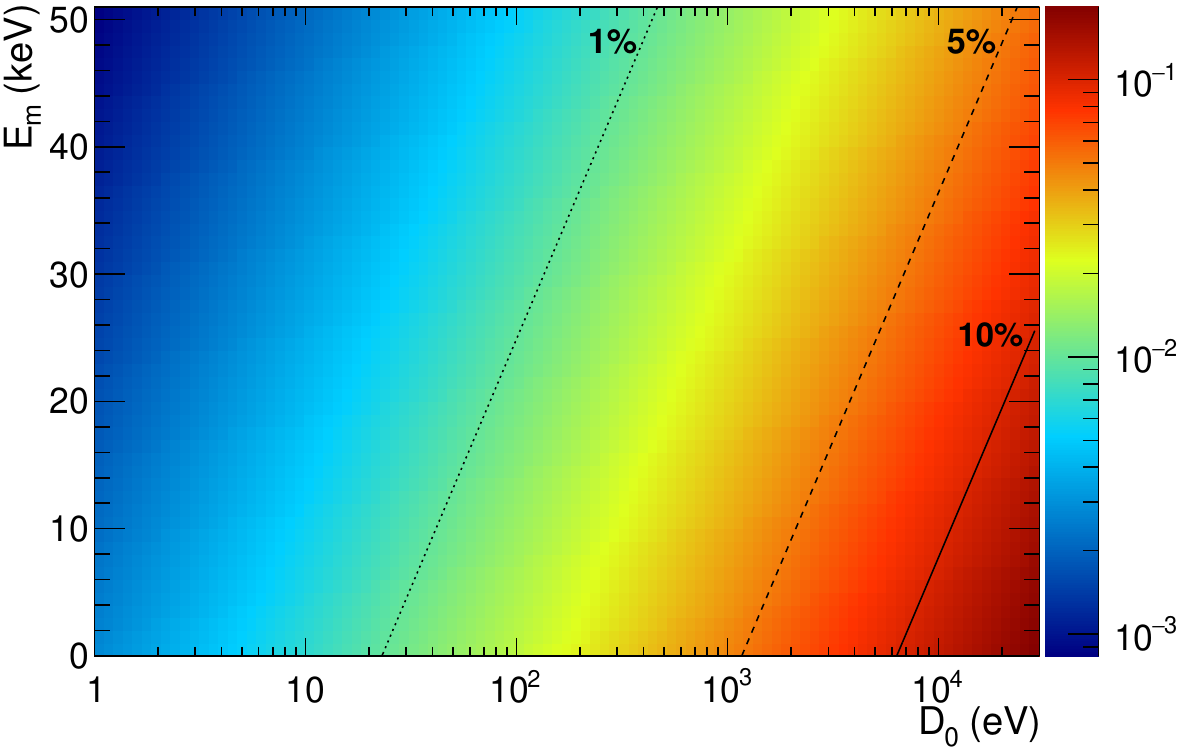}
\caption{$\delta_{MACS}^{E_m}$ as a function of $D_0$ and $E_m$ at $kT=30$~keV for Wigner NNS distribution.
}
\label{fig:kadonis1}
\end{figure}

Second, we determined relative uncertainty of MACS, $\delta_{\rm MACS}^{E_m}$, as the uncertainty of the MACS from energies $E_n>E_m$ with respect to total MACS. In this case $\delta_{\rm MACS}^{E_m}$ significantly decreases with $E_m$ as the contribution of the calculated part of the cross section to MACS decreases.
Considering Eq.~(\ref{eq:macs_areas}) approximately an exponential decrease with $E_m$ can be expected. Indeed, we found that for resonance sequences based on both GOE predictions and Wigner NNS the dependence can be well described by
\begin{equation}
\delta_{\rm MACS}^{E_m} = \delta_{\rm MACS} \times \exp \left( \frac{-1.25 E_m}{(kT)^{1.16}} \right).
\label{eq:MACS_dep}
\end{equation}
This dependence of $\delta_{\rm MACS}^{E_m}$ on $D_0$ and $E_m$ is then visualized in Fig.~\ref{fig:kadonis1} for $kT=30$ keV and Wigner NNS distribution; the figure for GOE and also $kT=8$~keV is available in the Suppl. Material. Although the actual $\delta_{\rm MACS}^{E_m}$ slightly depends on exact values of $\Gamma_\gamma$ and $S_\ell$ the formula reproduces simulated values for individual nuclei within a factor of about two. It means that once a reasonable estimate of $D_0$ is known, the relative uncertainty in MACS due to fluctuations can be estimated within this factor.

To get $\delta_{\rm MACS}$ for a real nucleus with partial measurements below $E_m$, the uncertainty from $E_n>E_m$ must then be combined with the experimental uncertainty from $E_n<E_m$. In practice, the knowledge of the fraction of MACS coming from $E_n>E_m$ is required for this calculation. We note that the presented values of $\delta_{\rm MACS}^{E_m}$ correspond to perfectly known average resonance parameters. These parameters are in reality always known only with a finite precision. 

%Description surely within a factor of 2 for all points but spacing 440 eV (Sr) and $E_m\gtrsim 15$ keV for GOE, where the difference is larger.

%%%%%%%%%%%%%%%%%%%%%%%%%%%%%%%%%%%%%%%%%%%%%%%%%%%%%%%%%%%%%%%%%
\section{Summary}

%{\it Should we stress anywhere that actual value of MACS can be in any case obtained only from the measurement? However, if we know average resonance parameters (precisely), we can use the result of this paper... } 

Even if all \emph{average} resonance parameters are precisely known the actual MACS can fluctuate around its expectation value. These fluctuations originate from different sources that may play various role depending on the values of average resonance parameters, exact fluctuation properties of level spacing, and the stellar temperature $kT$. In this contribution we analyzed impact of these individual sources. From that analysis, we derived  simple approximate formulae for relative uncertainty $\delta_{\rm MACS}$ of the MACS expectation value considering dependence on $kT$ and $s$-wave neutron resonance spacing $D_0$. Although there is also some impact of other average resonance parameters, presence of inelastic neutron scattering (or in general any open decay channel), or uncertainty in the spin distribution of resonance density, these effects are not dominant and they do not significantly change $\delta_{\rm MACS}$ with respect to the proposed dependences by more than about 50\%. 

The expression for estimating relative fluctuations partly depends on exact behavior of resonance positions. Many previous works, see, e.g. \cite{Haq82,Lombardi94} and references therein, indicated that actual resonance positions probably reasonably follow the prediction of the random matrix theory (GOE) that shows long-range correlations in individual resonance positions in addition to the Wigner distribution of the nearest-neighbor spacing. If this is really the case, the parametrization for the GOE sequences is the one to be used.
In any case, the relative fluctuation of MACS for $kT=8$ keV and 30 keV, relevant for $s$-process nucleosynthesis, is expected to be larger than about 10\% for $D_0$ above about one and six keV, respectively.

We further checked the impact of available experimental data on cross section at low neutron energies. Additional simple approximate formula was found to estimate the impact of unknown cross section only above a this energy on the possible value of MACS. 

Taken together, this work has identified the confidence that can be attributed to cross section calculations based on average resonance parameters for the MACS as a function of temperature and we provide clear metrics for the energy ranges where direct measurement would be needed in order to reduce the calculated cross section precision below a desired limit. While we have not considered any other cross section-weighting forms beyond a Maxwellian spectrum, this analysis generalizes trivially to other spectral shapes.

%{\bf Is there any required sentence on "data availability"?}

\begin{acknowledgments}
This work was supported by Grant No. 23-06439S of the Czech Science Foundation and by the US Department of Energy through the Los Alamos National Laboratory by the Laboratory Directed Research and Development program of Los Alamos National Laboratory under project number 20230052ER. Los Alamos National Laboratory is operated by Triad National Security, LLC, for the National Nuclear Security Administration of US Department of Energy (Contract No. 89233218CNA000001). 

\end{acknowledgments}

\bibliographystyle{apsrev4-2}
\bibliography{paper_pt,xsect_mk} 

\begin{thebibliography}{32}%
\makeatletter
\providecommand \@ifxundefined [1]{%
 \@ifx{#1\undefined}
}%
\providecommand \@ifnum [1]{%
 \ifnum #1\expandafter \@firstoftwo
 \else \expandafter \@secondoftwo
 \fi
}%
\providecommand \@ifx [1]{%
 \ifx #1\expandafter \@firstoftwo
 \else \expandafter \@secondoftwo
 \fi
}%
\providecommand \natexlab [1]{#1}%
\providecommand \enquote  [1]{``#1''}%
\providecommand \bibnamefont  [1]{#1}%
\providecommand \bibfnamefont [1]{#1}%
\providecommand \citenamefont [1]{#1}%
\providecommand \href@noop [0]{\@secondoftwo}%
\providecommand \href [0]{\begingroup \@sanitize@url \@href}%
\providecommand \@href[1]{\@@startlink{#1}\@@href}%
\providecommand \@@href[1]{\endgroup#1\@@endlink}%
\providecommand \@sanitize@url [0]{\catcode `\\12\catcode `\$12\catcode
  `\&12\catcode `\#12\catcode `\^12\catcode `\_12\catcode `\%12\relax}%
\providecommand \@@startlink[1]{}%
\providecommand \@@endlink[0]{}%
\providecommand \url  [0]{\begingroup\@sanitize@url \@url }%
\providecommand \@url [1]{\endgroup\@href {#1}{\urlprefix }}%
\providecommand \urlprefix  [0]{URL }%
\providecommand \Eprint [0]{\href }%
\providecommand \doibase [0]{https://doi.org/}%
\providecommand \selectlanguage [0]{\@gobble}%
\providecommand \bibinfo  [0]{\@secondoftwo}%
\providecommand \bibfield  [0]{\@secondoftwo}%
\providecommand \translation [1]{[#1]}%
\providecommand \BibitemOpen [0]{}%
\providecommand \bibitemStop [0]{}%
\providecommand \bibitemNoStop [0]{.\EOS\space}%
\providecommand \EOS [0]{\spacefactor3000\relax}%
\providecommand \BibitemShut  [1]{\csname bibitem#1\endcsname}%
\let\auto@bib@innerbib\@empty
%</preamble>
\bibitem [{\citenamefont {Burbidge}\ \emph {et~al.}(1957)\citenamefont
  {Burbidge}, \citenamefont {Burbidge}, \citenamefont {Fowler},\ and\
  \citenamefont {Hoyle}}]{BBF57}%
  \BibitemOpen
  \bibfield  {author} {\bibinfo {author} {\bibfnamefont {E.~M.}\ \bibnamefont
  {Burbidge}}, \bibinfo {author} {\bibfnamefont {G.~R.}\ \bibnamefont
  {Burbidge}}, \bibinfo {author} {\bibfnamefont {W.~A.}\ \bibnamefont
  {Fowler}},\ and\ \bibinfo {author} {\bibfnamefont {F.}~\bibnamefont
  {Hoyle}},\ }\href@noop {} {\bibfield  {journal} {\bibinfo  {journal} {Rev. of
  Mod. Phys.}\ }\textbf {\bibinfo {volume} {29}},\ \bibinfo {pages} {547}
  (\bibinfo {year} {1957})}\BibitemShut {NoStop}%
\bibitem [{\citenamefont {Wallerstein}\ \emph {et~al.}(1997)\citenamefont
  {Wallerstein}, \citenamefont {Iben}, \citenamefont {Parker}, \citenamefont
  {Boesgaard}, \citenamefont {Hale}, \citenamefont {Champagne}, \citenamefont
  {Barnes}, \citenamefont {K{\"a}ppeler}, \citenamefont {Smith}, \citenamefont
  {Hoffman}, \citenamefont {Timmes}, \citenamefont {Sneden}, \citenamefont
  {Boyd}, \citenamefont {Meyer},\ and\ \citenamefont {Lambert}}]{WIP97}%
  \BibitemOpen
  \bibfield  {author} {\bibinfo {author} {\bibfnamefont {G.}~\bibnamefont
  {Wallerstein}}, \bibinfo {author} {\bibfnamefont {I.}~\bibnamefont {Iben}},
  \bibinfo {author} {\bibfnamefont {P.}~\bibnamefont {Parker}}, \bibinfo
  {author} {\bibfnamefont {A.~M.}\ \bibnamefont {Boesgaard}}, \bibinfo {author}
  {\bibfnamefont {G.~M.}\ \bibnamefont {Hale}}, \bibinfo {author}
  {\bibfnamefont {A.~E.}\ \bibnamefont {Champagne}}, \bibinfo {author}
  {\bibfnamefont {C.~A.}\ \bibnamefont {Barnes}}, \bibinfo {author}
  {\bibfnamefont {F.}~\bibnamefont {K{\"a}ppeler}}, \bibinfo {author}
  {\bibfnamefont {V.~V.}\ \bibnamefont {Smith}}, \bibinfo {author}
  {\bibfnamefont {R.~D.}\ \bibnamefont {Hoffman}}, \bibinfo {author}
  {\bibfnamefont {F.~X.}\ \bibnamefont {Timmes}}, \bibinfo {author}
  {\bibfnamefont {C.}~\bibnamefont {Sneden}}, \bibinfo {author} {\bibfnamefont
  {R.~N.}\ \bibnamefont {Boyd}}, \bibinfo {author} {\bibfnamefont {B.~S.}\
  \bibnamefont {Meyer}},\ and\ \bibinfo {author} {\bibfnamefont {D.~L.}\
  \bibnamefont {Lambert}},\ }\href {https://doi.org/10.1103/RevModPhys.69.995}
  {\bibfield  {journal} {\bibinfo  {journal} {Rev. Mod. Phys.}\ }\textbf
  {\bibinfo {volume} {69}},\ \bibinfo {pages} {995} (\bibinfo {year}
  {1997})}\BibitemShut {NoStop}%
\bibitem [{\citenamefont {Fields}\ and\ \citenamefont {Wallner}(2023)}]{FiW23}%
  \BibitemOpen
  \bibfield  {author} {\bibinfo {author} {\bibfnamefont {B.~D.}\ \bibnamefont
  {Fields}}\ and\ \bibinfo {author} {\bibfnamefont {A.}~\bibnamefont
  {Wallner}},\ }\href
  {https://doi.org/https://doi.org/10.1146/annurev-nucl-011823-045541}
  {\bibfield  {journal} {\bibinfo  {journal} {Annual Review of Nuclear and
  Particle Science}\ }\textbf {\bibinfo {volume} {73}},\ \bibinfo {pages} {365}
  (\bibinfo {year} {2023})}\BibitemShut {NoStop}%
\bibitem [{\citenamefont {{Hayes}}\ \emph {et~al.}(2020)\citenamefont
  {{Hayes}}, \citenamefont {{Gooden}}, \citenamefont {{Henry}}, \citenamefont
  {{Jungman}}, \citenamefont {{Wilhelmy}}, \citenamefont {{Rundberg}},
  \citenamefont {{Yeamans}}, \citenamefont {{Kyrala}}, \citenamefont
  {{Cerjan}}, \citenamefont {{Danielson}}, \citenamefont {{Daligault}},
  \citenamefont {{Wilburn}}, \citenamefont {{Volegov}}, \citenamefont
  {{Wilde}}, \citenamefont {{Batha}}, \citenamefont {{Bredeweg}}, \citenamefont
  {{Kline}}, \citenamefont {{Grim}}, \citenamefont {{Hartouni}}, \citenamefont
  {{Shaughnessy}}, \citenamefont {{Velsko}}, \citenamefont {{Cassata}},
  \citenamefont {{Moody}}, \citenamefont {{Berzak Hopkins}}, \citenamefont
  {{Hinkel}}, \citenamefont {{D{\"o}ppner}}, \citenamefont {{Le Pape}},
  \citenamefont {{Graziani}}, \citenamefont {{Callahan}}, \citenamefont
  {{Hurricane}},\ and\ \citenamefont {{Schneider}}}]{HGH20}%
  \BibitemOpen
  \bibfield  {author} {\bibinfo {author} {\bibfnamefont {A.~C.}\ \bibnamefont
  {{Hayes}}}, \bibinfo {author} {\bibfnamefont {M.~E.}\ \bibnamefont
  {{Gooden}}}, \bibinfo {author} {\bibfnamefont {E.}~\bibnamefont {{Henry}}},
  \bibinfo {author} {\bibfnamefont {G.}~\bibnamefont {{Jungman}}}, \bibinfo
  {author} {\bibfnamefont {J.~B.}\ \bibnamefont {{Wilhelmy}}}, \bibinfo
  {author} {\bibfnamefont {R.~S.}\ \bibnamefont {{Rundberg}}}, \bibinfo
  {author} {\bibfnamefont {C.}~\bibnamefont {{Yeamans}}}, \bibinfo {author}
  {\bibfnamefont {G.}~\bibnamefont {{Kyrala}}}, \bibinfo {author}
  {\bibfnamefont {C.}~\bibnamefont {{Cerjan}}}, \bibinfo {author}
  {\bibfnamefont {D.~L.}\ \bibnamefont {{Danielson}}}, \bibinfo {author}
  {\bibfnamefont {J.}~\bibnamefont {{Daligault}}}, \bibinfo {author}
  {\bibfnamefont {C.}~\bibnamefont {{Wilburn}}}, \bibinfo {author}
  {\bibfnamefont {P.}~\bibnamefont {{Volegov}}}, \bibinfo {author}
  {\bibfnamefont {C.}~\bibnamefont {{Wilde}}}, \bibinfo {author} {\bibfnamefont
  {S.}~\bibnamefont {{Batha}}}, \bibinfo {author} {\bibfnamefont
  {T.}~\bibnamefont {{Bredeweg}}}, \bibinfo {author} {\bibfnamefont {J.~L.}\
  \bibnamefont {{Kline}}}, \bibinfo {author} {\bibfnamefont {G.~P.}\
  \bibnamefont {{Grim}}}, \bibinfo {author} {\bibfnamefont {E.~P.}\
  \bibnamefont {{Hartouni}}}, \bibinfo {author} {\bibfnamefont
  {D.}~\bibnamefont {{Shaughnessy}}}, \bibinfo {author} {\bibfnamefont
  {C.}~\bibnamefont {{Velsko}}}, \bibinfo {author} {\bibfnamefont {W.~S.}\
  \bibnamefont {{Cassata}}}, \bibinfo {author} {\bibfnamefont {K.}~\bibnamefont
  {{Moody}}}, \bibinfo {author} {\bibfnamefont {L.~F.}\ \bibnamefont {{Berzak
  Hopkins}}}, \bibinfo {author} {\bibfnamefont {D.}~\bibnamefont {{Hinkel}}},
  \bibinfo {author} {\bibfnamefont {T.}~\bibnamefont {{D{\"o}ppner}}}, \bibinfo
  {author} {\bibfnamefont {S.}~\bibnamefont {{Le Pape}}}, \bibinfo {author}
  {\bibfnamefont {F.}~\bibnamefont {{Graziani}}}, \bibinfo {author}
  {\bibfnamefont {D.~A.}\ \bibnamefont {{Callahan}}}, \bibinfo {author}
  {\bibfnamefont {O.~A.}\ \bibnamefont {{Hurricane}}},\ and\ \bibinfo {author}
  {\bibfnamefont {D.}~\bibnamefont {{Schneider}}},\ }\href
  {https://doi.org/10.1038/s41567-020-0790-3} {\bibfield  {journal} {\bibinfo
  {journal} {Nature Physics}\ }\textbf {\bibinfo {volume} {16}},\ \bibinfo
  {pages} {432} (\bibinfo {year} {2020})}\BibitemShut {NoStop}%
\bibitem [{\citenamefont {Aliotta}\ \emph {et~al.}(2021)\citenamefont
  {Aliotta}, \citenamefont {Buompane}, \citenamefont {Couder}, \citenamefont
  {Couture}, \citenamefont {deBoer}, \citenamefont {Formicola}, \citenamefont
  {Gialanella}, \citenamefont {Glorius}, \citenamefont {Imbriani},
  \citenamefont {Junker}, \citenamefont {Langer}, \citenamefont {Lennarz},
  \citenamefont {Litvinov}, \citenamefont {Liu}, \citenamefont {Lugaro},
  \citenamefont {Matei}, \citenamefont {Meisel}, \citenamefont {Piersanti},
  \citenamefont {Reifarth}, \citenamefont {Robertson}, \citenamefont {Simon},
  \citenamefont {Straniero}, \citenamefont {Tumino}, \citenamefont {Wiescher},\
  and\ \citenamefont {Xu}}]{ABC21}%
  \BibitemOpen
  \bibfield  {author} {\bibinfo {author} {\bibfnamefont {M.}~\bibnamefont
  {Aliotta}}, \bibinfo {author} {\bibfnamefont {R.}~\bibnamefont {Buompane}},
  \bibinfo {author} {\bibfnamefont {M.}~\bibnamefont {Couder}}, \bibinfo
  {author} {\bibfnamefont {A.}~\bibnamefont {Couture}}, \bibinfo {author}
  {\bibfnamefont {R.}~\bibnamefont {deBoer}}, \bibinfo {author} {\bibfnamefont
  {A.}~\bibnamefont {Formicola}}, \bibinfo {author} {\bibfnamefont
  {L.}~\bibnamefont {Gialanella}}, \bibinfo {author} {\bibfnamefont
  {J.}~\bibnamefont {Glorius}}, \bibinfo {author} {\bibfnamefont
  {G.}~\bibnamefont {Imbriani}}, \bibinfo {author} {\bibfnamefont
  {M.}~\bibnamefont {Junker}}, \bibinfo {author} {\bibfnamefont
  {C.}~\bibnamefont {Langer}}, \bibinfo {author} {\bibfnamefont
  {A.}~\bibnamefont {Lennarz}}, \bibinfo {author} {\bibfnamefont {Y.~A.}\
  \bibnamefont {Litvinov}}, \bibinfo {author} {\bibfnamefont {W.-P.}\
  \bibnamefont {Liu}}, \bibinfo {author} {\bibfnamefont {M.}~\bibnamefont
  {Lugaro}}, \bibinfo {author} {\bibfnamefont {C.}~\bibnamefont {Matei}},
  \bibinfo {author} {\bibfnamefont {Z.~P.}\ \bibnamefont {Meisel}}, \bibinfo
  {author} {\bibfnamefont {L.}~\bibnamefont {Piersanti}}, \bibinfo {author}
  {\bibfnamefont {R.}~\bibnamefont {Reifarth}}, \bibinfo {author}
  {\bibfnamefont {D.}~\bibnamefont {Robertson}}, \bibinfo {author}
  {\bibfnamefont {A.}~\bibnamefont {Simon}}, \bibinfo {author} {\bibfnamefont
  {O.}~\bibnamefont {Straniero}}, \bibinfo {author} {\bibfnamefont
  {A.}~\bibnamefont {Tumino}}, \bibinfo {author} {\bibfnamefont
  {M.}~\bibnamefont {Wiescher}},\ and\ \bibinfo {author} {\bibfnamefont
  {Y.}~\bibnamefont {Xu}},\ }\href
  {http://iopscience.iop.org/article/10.1088/1361-6471/ac2b0f} {\bibfield
  {journal} {\bibinfo  {journal} {Journal of Physics G: Nuclear and Particle
  Physics}\ } (\bibinfo {year} {2021})}\BibitemShut {NoStop}%
\bibitem [{\citenamefont {Koning}\ \emph {et~al.}(2006)\citenamefont {Koning},
  \citenamefont {Hilaire},\ and\ \citenamefont {Duijvestijn}}]{TALYS}%
  \BibitemOpen
  \bibfield  {author} {\bibinfo {author} {\bibfnamefont {A.}~\bibnamefont
  {Koning}}, \bibinfo {author} {\bibfnamefont {S.}~\bibnamefont {Hilaire}},\
  and\ \bibinfo {author} {\bibfnamefont {M.}~\bibnamefont {Duijvestijn}},\
  }\href {www.talys.eu} {\bibinfo {title} {Talys 1.8}} (\bibinfo {year}
  {2006}),\ \bibinfo {note} {www.talys.eu}\BibitemShut {NoStop}%
\bibitem [{\citenamefont {Kawano}(2024)}]{CoH}%
  \BibitemOpen
  \bibfield  {author} {\bibinfo {author} {\bibfnamefont {T.}~\bibnamefont
  {Kawano}},\ }\href {https://github.com/toshihikokawano/coh3} {\bibinfo
  {title} {Optical and statistical hauser-feshbach model code: Coh ver.3}}
  (\bibinfo {year} {2024})\BibitemShut {NoStop}%
\bibitem [{\citenamefont {Rochman}\ \emph {et~al.}(2017)\citenamefont
  {Rochman}, \citenamefont {Goriely}, \citenamefont {Koning},\ and\
  \citenamefont {Ferroukhi}}]{Rochman17}%
  \BibitemOpen
  \bibfield  {author} {\bibinfo {author} {\bibfnamefont {D.}~\bibnamefont
  {Rochman}}, \bibinfo {author} {\bibfnamefont {S.}~\bibnamefont {Goriely}},
  \bibinfo {author} {\bibfnamefont {A.~J.}\ \bibnamefont {Koning}},\ and\
  \bibinfo {author} {\bibfnamefont {H.}~\bibnamefont {Ferroukhi}},\ }\href
  {https://doi.org/10.1016/j.physletb.2016.11.018} {\bibfield  {journal}
  {\bibinfo  {journal} {Physics Letters B}\ }\textbf {\bibinfo {volume}
  {764}},\ \bibinfo {pages} {109} (\bibinfo {year} {2017})}\BibitemShut
  {NoStop}%
\bibitem [{\citenamefont {Beer}\ \emph {et~al.}(1992)\citenamefont {Beer},
  \citenamefont {Voss},\ and\ \citenamefont {Winters}}]{Beer92}%
  \BibitemOpen
  \bibfield  {author} {\bibinfo {author} {\bibfnamefont {H.}~\bibnamefont
  {Beer}}, \bibinfo {author} {\bibfnamefont {F.}~\bibnamefont {Voss}},\ and\
  \bibinfo {author} {\bibfnamefont {R.}~\bibnamefont {Winters}},\ }\href
  {https://doi.org/10.1086/191669} {\bibfield  {journal} {\bibinfo  {journal}
  {Astrophysical Journal Suppl. Series}\ }\textbf {\bibinfo {volume} {80}},\
  \bibinfo {pages} {403} (\bibinfo {year} {1992})}\BibitemShut {NoStop}%
\bibitem [{\citenamefont {von Egidy}\ and\ \citenamefont
  {Bucurescu}(2005)}]{Egidy05}%
  \BibitemOpen
  \bibfield  {author} {\bibinfo {author} {\bibfnamefont {T.}~\bibnamefont {von
  Egidy}}\ and\ \bibinfo {author} {\bibfnamefont {D.}~\bibnamefont
  {Bucurescu}},\ }\href {https://doi.org/10.1103/PhysRevC.72.044311} {\bibfield
   {journal} {\bibinfo  {journal} {Phys. Rev. C}\ }\textbf {\bibinfo {volume}
  {72}},\ \bibinfo {pages} {044311} (\bibinfo {year} {2005})},\ \bibinfo {note}
  {{\bf 73}, 049901(E) (2006)}\BibitemShut {NoStop}%
\bibitem [{\citenamefont {Fr\"{o}hner}(1992)}]{Frohner92}%
  \BibitemOpen
  \bibfield  {author} {\bibinfo {author} {\bibfnamefont {F.}~\bibnamefont
  {Fr\"{o}hner}},\ }\href@noop {} {\emph {\bibinfo {title} {Theory of Neutron
  Resonance Cross Sections for Safety Applications}}},\ \bibinfo {type} {Tech.
  Rep.}\ \bibinfo {number} {KfK 5073}\ (\bibinfo  {institution}
  {Kernforschungszentrum Karlsruhe GmbH, Karlsruhe},\ \bibinfo {year}
  {1992})\BibitemShut {NoStop}%
\bibitem [{\citenamefont {Lederer}\ \emph {et~al.}(2013)\citenamefont
  {Lederer}, \citenamefont {Massimi}, \citenamefont {Altstadt}, \citenamefont
  {Andrzejewski}, \citenamefont {Audouin}, \citenamefont {Barbagallo},
  \citenamefont {Becares}, \citenamefont {Becvar}, \citenamefont {Belloni},
  \citenamefont {Berthoumieux}, \citenamefont {Billowes}, \citenamefont
  {Boccone}, \citenamefont {Bosnar}, \citenamefont {Brugger}, \citenamefont
  {Calviani}, \citenamefont {Calvino}, \citenamefont {Cano-Ott}, \citenamefont
  {Carrapico}, \citenamefont {Cerutti}, \citenamefont {Chiaveri}, \citenamefont
  {Chin}, \citenamefont {Colonna}, \citenamefont {Cortes}, \citenamefont
  {Cortes-Giraldo}, \citenamefont {Diakaki}, \citenamefont {Domingo-Pardo},
  \citenamefont {Duran}, \citenamefont {Dressler}, \citenamefont {Dzysiuk},
  \citenamefont {Eleftheriadis}, \citenamefont {Ferrari}, \citenamefont
  {Fraval}, \citenamefont {Ganesan}, \citenamefont {Garcia}, \citenamefont
  {Giubrone}, \citenamefont {Gomez-Hornillos}, \citenamefont {Goncalves},
  \citenamefont {Gonzalez-Romero}, \citenamefont {Griesmayer}, \citenamefont
  {Guerrero}, \citenamefont {Gunsing}, \citenamefont {Gurusamy}, \citenamefont
  {Jenkins}, \citenamefont {Jericha}, \citenamefont {Kadi}, \citenamefont
  {Kaeppeler}, \citenamefont {Karadimos}, \citenamefont {Kivel}, \citenamefont
  {Koehler}, \citenamefont {Kokkoris}, \citenamefont {Korschinek},
  \citenamefont {Krticka}, \citenamefont {Kroll}, \citenamefont {Langer},
  \citenamefont {Leeb}, \citenamefont {Leong}, \citenamefont {Losito},
  \citenamefont {Manousos}, \citenamefont {Marganiec}, \citenamefont
  {Martinez}, \citenamefont {Mastinu}, \citenamefont {Mastromarco},
  \citenamefont {Meaze}, \citenamefont {Mendoza}, \citenamefont {Mengoni},
  \citenamefont {Milazzo}, \citenamefont {Mingrone}, \citenamefont {Mirea},
  \citenamefont {Mondelaers}, \citenamefont {Paradela}, \citenamefont {Pavlik},
  \citenamefont {Perkowski}, \citenamefont {Pignatari}, \citenamefont
  {Plompen}, \citenamefont {Praena}, \citenamefont {Quesada}, \citenamefont
  {Rauscher}, \citenamefont {Reifarth}, \citenamefont {Riego}, \citenamefont
  {Roman}, \citenamefont {Rubbia}, \citenamefont {Sarmento}, \citenamefont
  {Schillebeeckx}, \citenamefont {Schmidt}, \citenamefont {Schumann},
  \citenamefont {Tagliente}, \citenamefont {Tain}, \citenamefont {Tarrio},
  \citenamefont {Tassan-Got}, \citenamefont {Tsinganis}, \citenamefont
  {Valenta}, \citenamefont {Vannini}, \citenamefont {Variale}, \citenamefont
  {Vaz}, \citenamefont {Ventura}, \citenamefont {Versaci}, \citenamefont
  {Vermeulen}, \citenamefont {Vlachoudis}, \citenamefont {Vlastou},
  \citenamefont {Wallner}, \citenamefont {Ware}, \citenamefont {Weigand},
  \citenamefont {Weiss}, \citenamefont {Wright}, \citenamefont {Zugec},\ and\
  \citenamefont {n~TOF~Collaboration}}]{Lederer13}%
  \BibitemOpen
  \bibfield  {author} {\bibinfo {author} {\bibfnamefont {C.}~\bibnamefont
  {Lederer}}, \bibinfo {author} {\bibfnamefont {C.}~\bibnamefont {Massimi}},
  \bibinfo {author} {\bibfnamefont {S.}~\bibnamefont {Altstadt}}, \bibinfo
  {author} {\bibfnamefont {J.}~\bibnamefont {Andrzejewski}}, \bibinfo {author}
  {\bibfnamefont {L.}~\bibnamefont {Audouin}}, \bibinfo {author} {\bibfnamefont
  {M.}~\bibnamefont {Barbagallo}}, \bibinfo {author} {\bibfnamefont
  {V.}~\bibnamefont {Becares}}, \bibinfo {author} {\bibfnamefont
  {F.}~\bibnamefont {Becvar}}, \bibinfo {author} {\bibfnamefont
  {F.}~\bibnamefont {Belloni}}, \bibinfo {author} {\bibfnamefont
  {E.}~\bibnamefont {Berthoumieux}}, \bibinfo {author} {\bibfnamefont
  {J.}~\bibnamefont {Billowes}}, \bibinfo {author} {\bibfnamefont
  {V.}~\bibnamefont {Boccone}}, \bibinfo {author} {\bibfnamefont
  {D.}~\bibnamefont {Bosnar}}, \bibinfo {author} {\bibfnamefont
  {M.}~\bibnamefont {Brugger}}, \bibinfo {author} {\bibfnamefont
  {M.}~\bibnamefont {Calviani}}, \bibinfo {author} {\bibfnamefont
  {F.}~\bibnamefont {Calvino}}, \bibinfo {author} {\bibfnamefont
  {D.}~\bibnamefont {Cano-Ott}}, \bibinfo {author} {\bibfnamefont
  {C.}~\bibnamefont {Carrapico}}, \bibinfo {author} {\bibfnamefont
  {F.}~\bibnamefont {Cerutti}}, \bibinfo {author} {\bibfnamefont
  {E.}~\bibnamefont {Chiaveri}}, \bibinfo {author} {\bibfnamefont
  {M.}~\bibnamefont {Chin}}, \bibinfo {author} {\bibfnamefont {N.}~\bibnamefont
  {Colonna}}, \bibinfo {author} {\bibfnamefont {G.}~\bibnamefont {Cortes}},
  \bibinfo {author} {\bibfnamefont {M.~A.}\ \bibnamefont {Cortes-Giraldo}},
  \bibinfo {author} {\bibfnamefont {M.}~\bibnamefont {Diakaki}}, \bibinfo
  {author} {\bibfnamefont {C.}~\bibnamefont {Domingo-Pardo}}, \bibinfo {author}
  {\bibfnamefont {I.}~\bibnamefont {Duran}}, \bibinfo {author} {\bibfnamefont
  {R.}~\bibnamefont {Dressler}}, \bibinfo {author} {\bibfnamefont
  {N.}~\bibnamefont {Dzysiuk}}, \bibinfo {author} {\bibfnamefont
  {C.}~\bibnamefont {Eleftheriadis}}, \bibinfo {author} {\bibfnamefont
  {A.}~\bibnamefont {Ferrari}}, \bibinfo {author} {\bibfnamefont
  {K.}~\bibnamefont {Fraval}}, \bibinfo {author} {\bibfnamefont
  {S.}~\bibnamefont {Ganesan}}, \bibinfo {author} {\bibfnamefont {A.~R.}\
  \bibnamefont {Garcia}}, \bibinfo {author} {\bibfnamefont {G.}~\bibnamefont
  {Giubrone}}, \bibinfo {author} {\bibfnamefont {M.~B.}\ \bibnamefont
  {Gomez-Hornillos}}, \bibinfo {author} {\bibfnamefont {I.~F.}\ \bibnamefont
  {Goncalves}}, \bibinfo {author} {\bibfnamefont {E.}~\bibnamefont
  {Gonzalez-Romero}}, \bibinfo {author} {\bibfnamefont {E.}~\bibnamefont
  {Griesmayer}}, \bibinfo {author} {\bibfnamefont {C.}~\bibnamefont
  {Guerrero}}, \bibinfo {author} {\bibfnamefont {F.}~\bibnamefont {Gunsing}},
  \bibinfo {author} {\bibfnamefont {P.}~\bibnamefont {Gurusamy}}, \bibinfo
  {author} {\bibfnamefont {D.~G.}\ \bibnamefont {Jenkins}}, \bibinfo {author}
  {\bibfnamefont {E.}~\bibnamefont {Jericha}}, \bibinfo {author} {\bibfnamefont
  {Y.}~\bibnamefont {Kadi}}, \bibinfo {author} {\bibfnamefont {F.}~\bibnamefont
  {Kaeppeler}}, \bibinfo {author} {\bibfnamefont {D.}~\bibnamefont
  {Karadimos}}, \bibinfo {author} {\bibfnamefont {N.}~\bibnamefont {Kivel}},
  \bibinfo {author} {\bibfnamefont {P.}~\bibnamefont {Koehler}}, \bibinfo
  {author} {\bibfnamefont {M.}~\bibnamefont {Kokkoris}}, \bibinfo {author}
  {\bibfnamefont {G.}~\bibnamefont {Korschinek}}, \bibinfo {author}
  {\bibfnamefont {M.}~\bibnamefont {Krticka}}, \bibinfo {author} {\bibfnamefont
  {J.}~\bibnamefont {Kroll}}, \bibinfo {author} {\bibfnamefont
  {C.}~\bibnamefont {Langer}}, \bibinfo {author} {\bibfnamefont
  {H.}~\bibnamefont {Leeb}}, \bibinfo {author} {\bibfnamefont {L.~S.}\
  \bibnamefont {Leong}}, \bibinfo {author} {\bibfnamefont {R.}~\bibnamefont
  {Losito}}, \bibinfo {author} {\bibfnamefont {A.}~\bibnamefont {Manousos}},
  \bibinfo {author} {\bibfnamefont {J.}~\bibnamefont {Marganiec}}, \bibinfo
  {author} {\bibfnamefont {T.}~\bibnamefont {Martinez}}, \bibinfo {author}
  {\bibfnamefont {P.~F.}\ \bibnamefont {Mastinu}}, \bibinfo {author}
  {\bibfnamefont {M.}~\bibnamefont {Mastromarco}}, \bibinfo {author}
  {\bibfnamefont {M.}~\bibnamefont {Meaze}}, \bibinfo {author} {\bibfnamefont
  {E.}~\bibnamefont {Mendoza}}, \bibinfo {author} {\bibfnamefont
  {A.}~\bibnamefont {Mengoni}}, \bibinfo {author} {\bibfnamefont {P.~M.}\
  \bibnamefont {Milazzo}}, \bibinfo {author} {\bibfnamefont {F.}~\bibnamefont
  {Mingrone}}, \bibinfo {author} {\bibfnamefont {M.}~\bibnamefont {Mirea}},
  \bibinfo {author} {\bibfnamefont {W.}~\bibnamefont {Mondelaers}}, \bibinfo
  {author} {\bibfnamefont {C.}~\bibnamefont {Paradela}}, \bibinfo {author}
  {\bibfnamefont {A.}~\bibnamefont {Pavlik}}, \bibinfo {author} {\bibfnamefont
  {J.}~\bibnamefont {Perkowski}}, \bibinfo {author} {\bibfnamefont
  {M.}~\bibnamefont {Pignatari}}, \bibinfo {author} {\bibfnamefont
  {A.}~\bibnamefont {Plompen}}, \bibinfo {author} {\bibfnamefont
  {J.}~\bibnamefont {Praena}}, \bibinfo {author} {\bibfnamefont {J.~M.}\
  \bibnamefont {Quesada}}, \bibinfo {author} {\bibfnamefont {T.}~\bibnamefont
  {Rauscher}}, \bibinfo {author} {\bibfnamefont {R.}~\bibnamefont {Reifarth}},
  \bibinfo {author} {\bibfnamefont {A.}~\bibnamefont {Riego}}, \bibinfo
  {author} {\bibfnamefont {F.}~\bibnamefont {Roman}}, \bibinfo {author}
  {\bibfnamefont {C.}~\bibnamefont {Rubbia}}, \bibinfo {author} {\bibfnamefont
  {R.}~\bibnamefont {Sarmento}}, \bibinfo {author} {\bibfnamefont
  {P.}~\bibnamefont {Schillebeeckx}}, \bibinfo {author} {\bibfnamefont
  {S.}~\bibnamefont {Schmidt}}, \bibinfo {author} {\bibfnamefont
  {D.}~\bibnamefont {Schumann}}, \bibinfo {author} {\bibfnamefont
  {G.}~\bibnamefont {Tagliente}}, \bibinfo {author} {\bibfnamefont {J.~L.}\
  \bibnamefont {Tain}}, \bibinfo {author} {\bibfnamefont {D.}~\bibnamefont
  {Tarrio}}, \bibinfo {author} {\bibfnamefont {L.}~\bibnamefont {Tassan-Got}},
  \bibinfo {author} {\bibfnamefont {A.}~\bibnamefont {Tsinganis}}, \bibinfo
  {author} {\bibfnamefont {S.}~\bibnamefont {Valenta}}, \bibinfo {author}
  {\bibfnamefont {G.}~\bibnamefont {Vannini}}, \bibinfo {author} {\bibfnamefont
  {V.}~\bibnamefont {Variale}}, \bibinfo {author} {\bibfnamefont
  {P.}~\bibnamefont {Vaz}}, \bibinfo {author} {\bibfnamefont {A.}~\bibnamefont
  {Ventura}}, \bibinfo {author} {\bibfnamefont {R.}~\bibnamefont {Versaci}},
  \bibinfo {author} {\bibfnamefont {M.~J.}\ \bibnamefont {Vermeulen}}, \bibinfo
  {author} {\bibfnamefont {V.}~\bibnamefont {Vlachoudis}}, \bibinfo {author}
  {\bibfnamefont {R.}~\bibnamefont {Vlastou}}, \bibinfo {author} {\bibfnamefont
  {A.}~\bibnamefont {Wallner}}, \bibinfo {author} {\bibfnamefont
  {T.}~\bibnamefont {Ware}}, \bibinfo {author} {\bibfnamefont {M.}~\bibnamefont
  {Weigand}}, \bibinfo {author} {\bibfnamefont {C.}~\bibnamefont {Weiss}},
  \bibinfo {author} {\bibfnamefont {T.~J.}\ \bibnamefont {Wright}}, \bibinfo
  {author} {\bibfnamefont {P.}~\bibnamefont {Zugec}},\ and\ \bibinfo {author}
  {\bibnamefont {n~TOF~Collaboration}},\ }\bibfield  {journal} {\bibinfo
  {journal} {Physical Review Letters}\ }\textbf {\bibinfo {volume} {110}},\
  \href {https://doi.org/10.1103/PhysRevLett.110.022501}
  {10.1103/PhysRevLett.110.022501} (\bibinfo {year} {2013})\BibitemShut
  {NoStop}%
\bibitem [{\citenamefont {Weigand}\ \emph {et~al.}(2015)\citenamefont
  {Weigand}, \citenamefont {Bredeweg}, \citenamefont {Couture}, \citenamefont
  {Goebel}, \citenamefont {Heftrich}, \citenamefont {Jandel}, \citenamefont
  {Kaeppeler}, \citenamefont {Lederer}, \citenamefont {Kivel}, \citenamefont
  {Korschinek}, \citenamefont {Krticka}, \citenamefont {O'Donnell},
  \citenamefont {Ostermoeller}, \citenamefont {Plag}, \citenamefont {Reifarth},
  \citenamefont {Schumann}, \citenamefont {Ullmann},\ and\ \citenamefont
  {Wallner}}]{Weigand15}%
  \BibitemOpen
  \bibfield  {author} {\bibinfo {author} {\bibfnamefont {M.}~\bibnamefont
  {Weigand}}, \bibinfo {author} {\bibfnamefont {T.~A.}\ \bibnamefont
  {Bredeweg}}, \bibinfo {author} {\bibfnamefont {A.}~\bibnamefont {Couture}},
  \bibinfo {author} {\bibfnamefont {K.}~\bibnamefont {Goebel}}, \bibinfo
  {author} {\bibfnamefont {T.}~\bibnamefont {Heftrich}}, \bibinfo {author}
  {\bibfnamefont {M.}~\bibnamefont {Jandel}}, \bibinfo {author} {\bibfnamefont
  {F.}~\bibnamefont {Kaeppeler}}, \bibinfo {author} {\bibfnamefont
  {C.}~\bibnamefont {Lederer}}, \bibinfo {author} {\bibfnamefont
  {N.}~\bibnamefont {Kivel}}, \bibinfo {author} {\bibfnamefont
  {G.}~\bibnamefont {Korschinek}}, \bibinfo {author} {\bibfnamefont
  {M.}~\bibnamefont {Krticka}}, \bibinfo {author} {\bibfnamefont {J.~M.}\
  \bibnamefont {O'Donnell}}, \bibinfo {author} {\bibfnamefont {J.}~\bibnamefont
  {Ostermoeller}}, \bibinfo {author} {\bibfnamefont {R.}~\bibnamefont {Plag}},
  \bibinfo {author} {\bibfnamefont {R.}~\bibnamefont {Reifarth}}, \bibinfo
  {author} {\bibfnamefont {D.}~\bibnamefont {Schumann}}, \bibinfo {author}
  {\bibfnamefont {J.~L.}\ \bibnamefont {Ullmann}},\ and\ \bibinfo {author}
  {\bibfnamefont {A.}~\bibnamefont {Wallner}},\ }\bibfield  {journal} {\bibinfo
   {journal} {Physical Review C}\ }\textbf {\bibinfo {volume} {92}},\ \href
  {https://doi.org/10.1103/PhysRevC.92.045810} {10.1103/PhysRevC.92.045810}
  (\bibinfo {year} {2015})\BibitemShut {NoStop}%
\bibitem [{\citenamefont {Porter}\ and\ \citenamefont
  {Thomas}(1956)}]{Porter56}%
  \BibitemOpen
  \bibfield  {author} {\bibinfo {author} {\bibfnamefont {C.~E.}\ \bibnamefont
  {Porter}}\ and\ \bibinfo {author} {\bibfnamefont {R.~G.}\ \bibnamefont
  {Thomas}},\ }\href {https://doi.org/10.1103/PhysRev.104.483} {\bibfield
  {journal} {\bibinfo  {journal} {Phys. Rev.}\ }\textbf {\bibinfo {volume}
  {104}},\ \bibinfo {pages} {483} (\bibinfo {year} {1956})}\BibitemShut
  {NoStop}%
\bibitem [{\citenamefont {K\"appeler}\ \emph {et~al.}(2011)\citenamefont
  {K\"appeler}, \citenamefont {Gallino}, \citenamefont {Bisterzo},\ and\
  \citenamefont {Aoki}}]{KGB11}%
  \BibitemOpen
  \bibfield  {author} {\bibinfo {author} {\bibfnamefont {F.}~\bibnamefont
  {K\"appeler}}, \bibinfo {author} {\bibfnamefont {R.}~\bibnamefont {Gallino}},
  \bibinfo {author} {\bibfnamefont {S.}~\bibnamefont {Bisterzo}},\ and\
  \bibinfo {author} {\bibfnamefont {W.}~\bibnamefont {Aoki}},\ }\href
  {https://doi.org/10.1103/RevModPhys.83.157} {\bibfield  {journal} {\bibinfo
  {journal} {Rev. Mod. Phys.}\ }\textbf {\bibinfo {volume} {83}},\ \bibinfo
  {pages} {157} (\bibinfo {year} {2011})}\BibitemShut {NoStop}%
\bibitem [{\citenamefont {Be\v{c}v\'{a}\v{r}}(1998)}]{Becvar98}%
  \BibitemOpen
  \bibfield  {author} {\bibinfo {author} {\bibfnamefont {F.}~\bibnamefont
  {Be\v{c}v\'{a}\v{r}}},\ }\href
  {https://doi.org/10.1016/S0168-9002(98)00787-6} {\bibfield  {journal}
  {\bibinfo  {journal} {Nucl. Instrum. Methods A}\ }\textbf {\bibinfo {volume}
  {417}},\ \bibinfo {pages} {434} (\bibinfo {year} {1998})}\BibitemShut
  {NoStop}%
\bibitem [{\citenamefont {Mughabghab}(2018{\natexlab{a}})}]{Mughabghab18a}%
  \BibitemOpen
  \bibfield  {author} {\bibinfo {author} {\bibfnamefont {S.~F.}\ \bibnamefont
  {Mughabghab}},\ }\href {https://doi.org/10.1016/C2015-0-00522-6} {\emph
  {\bibinfo {title} {{Atlas of Neutron Resonances}}}}\ (\bibinfo  {publisher}
  {Elsevier},\ \bibinfo {address} {Amsterdam},\ \bibinfo {year} {2018})\
  \bibinfo {note} {{Resonance Properties and Thermal Cross Sections
  Z=1-60}}\BibitemShut {NoStop}%
\bibitem [{\citenamefont {Mughabghab}(2018{\natexlab{b}})}]{Mughabghab18b}%
  \BibitemOpen
  \bibfield  {author} {\bibinfo {author} {\bibfnamefont {S.~F.}\ \bibnamefont
  {Mughabghab}},\ }\href {https://doi.org/10.1016/C2015-0-00524-X} {\emph
  {\bibinfo {title} {{Atlas of Neutron Resonances}}}}\ (\bibinfo  {publisher}
  {Elsevier},\ \bibinfo {address} {Amsterdam},\ \bibinfo {year} {2018})\
  \bibinfo {note} {{Resonance Properties and Thermal Cross Sections
  Z=61-102}}\BibitemShut {NoStop}%
\bibitem [{\citenamefont {Capote}\ \emph {et~al.}(2009)\citenamefont {Capote},
  \citenamefont {Herman}, \citenamefont {Oblozinsky}, \citenamefont {Young},
  \citenamefont {Goriely}, \citenamefont {Belgya}, \citenamefont {Ignatyuk},
  \citenamefont {Koning}, \citenamefont {Hilaire}, \citenamefont {Plujko},
  \citenamefont {Avrigeanu}, \citenamefont {Bersillon}, \citenamefont
  {Chadwick}, \citenamefont {Fukahori}, \citenamefont {Ge}, \citenamefont
  {Han}, \citenamefont {Kailas}, \citenamefont {Kopecky}, \citenamefont
  {Maslov}, \citenamefont {Reffo}, \citenamefont {Sin}, \citenamefont
  {Soukhovitskii},\ and\ \citenamefont {Talou}}]{RIPL3}%
  \BibitemOpen
  \bibfield  {author} {\bibinfo {author} {\bibfnamefont {R.}~\bibnamefont
  {Capote}}, \bibinfo {author} {\bibfnamefont {M.}~\bibnamefont {Herman}},
  \bibinfo {author} {\bibfnamefont {P.}~\bibnamefont {Oblozinsky}}, \bibinfo
  {author} {\bibfnamefont {P.}~\bibnamefont {Young}}, \bibinfo {author}
  {\bibfnamefont {S.}~\bibnamefont {Goriely}}, \bibinfo {author} {\bibfnamefont
  {T.}~\bibnamefont {Belgya}}, \bibinfo {author} {\bibfnamefont
  {A.}~\bibnamefont {Ignatyuk}}, \bibinfo {author} {\bibfnamefont
  {A.}~\bibnamefont {Koning}}, \bibinfo {author} {\bibfnamefont
  {S.}~\bibnamefont {Hilaire}}, \bibinfo {author} {\bibfnamefont
  {V.}~\bibnamefont {Plujko}}, \bibinfo {author} {\bibfnamefont
  {M.}~\bibnamefont {Avrigeanu}}, \bibinfo {author} {\bibfnamefont
  {O.}~\bibnamefont {Bersillon}}, \bibinfo {author} {\bibfnamefont
  {M.}~\bibnamefont {Chadwick}}, \bibinfo {author} {\bibfnamefont
  {T.}~\bibnamefont {Fukahori}}, \bibinfo {author} {\bibfnamefont
  {Z.}~\bibnamefont {Ge}}, \bibinfo {author} {\bibfnamefont {Y.}~\bibnamefont
  {Han}}, \bibinfo {author} {\bibfnamefont {S.}~\bibnamefont {Kailas}},
  \bibinfo {author} {\bibfnamefont {J.}~\bibnamefont {Kopecky}}, \bibinfo
  {author} {\bibfnamefont {V.}~\bibnamefont {Maslov}}, \bibinfo {author}
  {\bibfnamefont {G.}~\bibnamefont {Reffo}}, \bibinfo {author} {\bibfnamefont
  {M.}~\bibnamefont {Sin}}, \bibinfo {author} {\bibfnamefont {E.}~\bibnamefont
  {Soukhovitskii}},\ and\ \bibinfo {author} {\bibfnamefont {P.}~\bibnamefont
  {Talou}},\ }\href {https://doi.org/10.1016/j.nds.2009.10.004} {\bibfield
  {journal} {\bibinfo  {journal} {Nucl. Data Sheets}\ }\textbf {\bibinfo
  {volume} {110}},\ \bibinfo {pages} {3107} (\bibinfo {year}
  {2009})}\BibitemShut {NoStop}%
\bibitem [{\citenamefont {Wigner}(1959)}]{Wigner59}%
  \BibitemOpen
  \bibfield  {author} {\bibinfo {author} {\bibfnamefont {E.}~\bibnamefont
  {Wigner}},\ }in\ \href@noop {} {\emph {\bibinfo {booktitle} {Proc. Conf.
  Appl. Math.}}}\ (\bibinfo {year} {1959})\ p.\ \bibinfo {pages}
  {483}\BibitemShut {NoStop}%
\bibitem [{\citenamefont {Rochman}\ \emph {et~al.}(2013)\citenamefont
  {Rochman}, \citenamefont {Koning}, \citenamefont {Kopecky}, \citenamefont
  {Sublet}, \citenamefont {Ribon},\ and\ \citenamefont {Moxon}}]{Rochman13}%
  \BibitemOpen
  \bibfield  {author} {\bibinfo {author} {\bibfnamefont {D.}~\bibnamefont
  {Rochman}}, \bibinfo {author} {\bibfnamefont {A.~J.}\ \bibnamefont {Koning}},
  \bibinfo {author} {\bibfnamefont {J.}~\bibnamefont {Kopecky}}, \bibinfo
  {author} {\bibfnamefont {J.~C.}\ \bibnamefont {Sublet}}, \bibinfo {author}
  {\bibfnamefont {P.}~\bibnamefont {Ribon}},\ and\ \bibinfo {author}
  {\bibfnamefont {M.}~\bibnamefont {Moxon}},\ }\href
  {https://doi.org/10.1016/j.anucene.2012.08.015} {\bibfield  {journal}
  {\bibinfo  {journal} {Annals of Nuclear Energy}\ }\textbf {\bibinfo {volume}
  {51}},\ \bibinfo {pages} {60} (\bibinfo {year} {2013})}\BibitemShut {NoStop}%
\bibitem [{\citenamefont {Dyson}\ and\ \citenamefont {Mehta}(1963)}]{Dyson63}%
  \BibitemOpen
  \bibfield  {author} {\bibinfo {author} {\bibfnamefont {F.~J.}\ \bibnamefont
  {Dyson}}\ and\ \bibinfo {author} {\bibfnamefont {M.~L.}\ \bibnamefont
  {Mehta}},\ }\href {https://doi.org/10.1063/1.1704008} {\bibfield  {journal}
  {\bibinfo  {journal} {Journal of Mathematical Physics}\ }\textbf {\bibinfo
  {volume} {4}},\ \bibinfo {pages} {701} (\bibinfo {year} {1963})}\BibitemShut
  {NoStop}%
\bibitem [{\citenamefont {Haq}\ \emph {et~al.}(1982)\citenamefont {Haq},
  \citenamefont {Pandey},\ and\ \citenamefont {Bohigas}}]{Haq82}%
  \BibitemOpen
  \bibfield  {author} {\bibinfo {author} {\bibfnamefont {R.}~\bibnamefont
  {Haq}}, \bibinfo {author} {\bibfnamefont {A.}~\bibnamefont {Pandey}},\ and\
  \bibinfo {author} {\bibfnamefont {O.}~\bibnamefont {Bohigas}},\ }\href
  {https://doi.org/10.1103/PhysRevLett.48.1086} {\bibfield  {journal} {\bibinfo
   {journal} {Physical Review Letters}\ }\textbf {\bibinfo {volume} {48}},\
  \bibinfo {pages} {1086} (\bibinfo {year} {1982})}\BibitemShut {NoStop}%
\bibitem [{\citenamefont {Koehler}\ \emph {et~al.}(2013)\citenamefont
  {Koehler}, \citenamefont {Becvar}, \citenamefont {Krticka}, \citenamefont
  {Guber},\ and\ \citenamefont {Ullmann}}]{Koehler13}%
  \BibitemOpen
  \bibfield  {author} {\bibinfo {author} {\bibfnamefont {P.~E.}\ \bibnamefont
  {Koehler}}, \bibinfo {author} {\bibfnamefont {F.}~\bibnamefont {Becvar}},
  \bibinfo {author} {\bibfnamefont {M.}~\bibnamefont {Krticka}}, \bibinfo
  {author} {\bibfnamefont {K.~H.}\ \bibnamefont {Guber}},\ and\ \bibinfo
  {author} {\bibfnamefont {J.~L.}\ \bibnamefont {Ullmann}},\ }\href
  {https://doi.org/10.1002/prop.201200067} {\bibfield  {journal} {\bibinfo
  {journal} {Fortschritte der Physik - Progress of Physics}\ }\textbf {\bibinfo
  {volume} {61}},\ \bibinfo {pages} {80} (\bibinfo {year} {2013})}\BibitemShut
  {NoStop}%
\bibitem [{\citenamefont {Krti\v{c}ka}\ and\ \citenamefont
  {Couture}()}]{SupplMat}%
  \BibitemOpen
  \bibfield  {author} {\bibinfo {author} {\bibfnamefont {M.}~\bibnamefont
  {Krti\v{c}ka}}\ and\ \bibinfo {author} {\bibfnamefont {A.}~\bibnamefont
  {Couture}},\ }\href@noop {} {\bibinfo {title} {Supplemental material}},\
  \bibinfo {howpublished} {\url{https://...}},\ \bibinfo {note}
  {\url{https://...}}\BibitemShut {Stop}%
\bibitem [{\citenamefont {WAGONER}(1969)}]{Wagoner69}%
  \BibitemOpen
  \bibfield  {author} {\bibinfo {author} {\bibfnamefont {R.}~\bibnamefont
  {WAGONER}},\ }\href {https://doi.org/10.1086/190191} {\bibfield  {journal}
  {\bibinfo  {journal} {ASTROPHYSICAL JOURNAL SUPPLEMENT SERIES}\ }\textbf
  {\bibinfo {volume} {18}},\ \bibinfo {pages} {247} (\bibinfo {year}
  {1969})}\BibitemShut {NoStop}%
\bibitem [{\citenamefont {Rauscher}\ \emph {et~al.}(1997)\citenamefont
  {Rauscher}, \citenamefont {Thielemann},\ and\ \citenamefont
  {Kratz}}]{Rauscher97}%
  \BibitemOpen
  \bibfield  {author} {\bibinfo {author} {\bibfnamefont {T.}~\bibnamefont
  {Rauscher}}, \bibinfo {author} {\bibfnamefont {F.}~\bibnamefont
  {Thielemann}},\ and\ \bibinfo {author} {\bibfnamefont {K.}~\bibnamefont
  {Kratz}},\ }\href {https://doi.org/10.1103/PhysRevC.56.1613} {\bibfield
  {journal} {\bibinfo  {journal} {Physical Review C}\ }\textbf {\bibinfo
  {volume} {56}},\ \bibinfo {pages} {1613} (\bibinfo {year}
  {1997})}\BibitemShut {NoStop}%
\bibitem [{\citenamefont {von Egidy}\ and\ \citenamefont
  {Bucurescu}(2009)}]{Egidy09}%
  \BibitemOpen
  \bibfield  {author} {\bibinfo {author} {\bibfnamefont {T.}~\bibnamefont {von
  Egidy}}\ and\ \bibinfo {author} {\bibfnamefont {D.}~\bibnamefont
  {Bucurescu}},\ }\href {https://doi.org/10.1103/PhysRevC.80.054310} {\bibfield
   {journal} {\bibinfo  {journal} {Phys. Rev. C}\ }\textbf {\bibinfo {volume}
  {80}},\ \bibinfo {pages} {054310} (\bibinfo {year} {2009})}\BibitemShut
  {NoStop}%
\bibitem [{\citenamefont {Dillmann}\ \emph {et~al.}(2006)\citenamefont
  {Dillmann}, \citenamefont {Heil}, \citenamefont {K\"{a}ppeler}, \citenamefont
  {Plag}, \citenamefont {Rauscher},\ and\ \citenamefont
  {Thielemann}}]{Kadonis}%
  \BibitemOpen
  \bibfield  {author} {\bibinfo {author} {\bibfnamefont {I.}~\bibnamefont
  {Dillmann}}, \bibinfo {author} {\bibfnamefont {M.}~\bibnamefont {Heil}},
  \bibinfo {author} {\bibfnamefont {F.}~\bibnamefont {K\"{a}ppeler}}, \bibinfo
  {author} {\bibfnamefont {R.}~\bibnamefont {Plag}}, \bibinfo {author}
  {\bibfnamefont {T.}~\bibnamefont {Rauscher}},\ and\ \bibinfo {author}
  {\bibfnamefont {F.}~\bibnamefont {Thielemann}},\ }in\ \href@noop {} {\emph
  {\bibinfo {booktitle} {AIP Conf. Proc.}}},\ Vol.\ \bibinfo {volume} {819}\
  (\bibinfo {year} {2006})\ p.\ \bibinfo {pages} {123},\ \bibinfo {note}
  {\url{https://www.kadonis.org/}}\BibitemShut {NoStop}%
\bibitem [{\citenamefont {Prokop}\ \emph {et~al.}(2019)\citenamefont {Prokop},
  \citenamefont {Couture}, \citenamefont {Jones}, \citenamefont {Mosby},
  \citenamefont {Rusev}, \citenamefont {Ullmann},\ and\ \citenamefont
  {Krticka}}]{Prokop19}%
  \BibitemOpen
  \bibfield  {author} {\bibinfo {author} {\bibfnamefont {C.~J.}\ \bibnamefont
  {Prokop}}, \bibinfo {author} {\bibfnamefont {A.}~\bibnamefont {Couture}},
  \bibinfo {author} {\bibfnamefont {S.}~\bibnamefont {Jones}}, \bibinfo
  {author} {\bibfnamefont {S.}~\bibnamefont {Mosby}}, \bibinfo {author}
  {\bibfnamefont {G.}~\bibnamefont {Rusev}}, \bibinfo {author} {\bibfnamefont
  {J.}~\bibnamefont {Ullmann}},\ and\ \bibinfo {author} {\bibfnamefont
  {M.}~\bibnamefont {Krticka}},\ }\bibfield  {journal} {\bibinfo  {journal}
  {Physical Review C}\ }\textbf {\bibinfo {volume} {99}},\ \href
  {https://doi.org/10.1103/PhysRevC.99.055809} {10.1103/PhysRevC.99.055809}
  (\bibinfo {year} {2019})\BibitemShut {NoStop}%
\bibitem [{\citenamefont {Knapová}\ and\ \citenamefont
  {Valenta}(2025)}]{Knapova25}%
  \BibitemOpen
  \bibfield  {author} {\bibinfo {author} {\bibfnamefont {H.~K. C. A. F. C. G.
  F. K. T. K. K. J. K. M. C. E. L. P. C. J. R. R. R. G. U. J.~L.}\ \bibnamefont
  {Knapová}, \bibfnamefont {I.}}\ and\ \bibinfo {author} {\bibfnamefont
  {S.}~\bibnamefont {Valenta}},\ }\href {https://doi.org/10.1103/5sjr-vyjv}
  {\bibfield  {journal} {\bibinfo  {journal} {Phys. Rev. C}\ ,\ } (\bibinfo
  {year} {2025})}\BibitemShut {NoStop}%
\bibitem [{\citenamefont {Lombardi}\ \emph {et~al.}(1994)\citenamefont
  {Lombardi}, \citenamefont {Bohigas},\ and\ \citenamefont
  {Seligman}}]{Lombardi94}%
  \BibitemOpen
  \bibfield  {author} {\bibinfo {author} {\bibfnamefont {M.}~\bibnamefont
  {Lombardi}}, \bibinfo {author} {\bibfnamefont {O.}~\bibnamefont {Bohigas}},\
  and\ \bibinfo {author} {\bibfnamefont {T.}~\bibnamefont {Seligman}},\ }\href
  {https://doi.org/10.1016/0370-2693(94)90191-0} {\bibfield  {journal}
  {\bibinfo  {journal} {Physics Letters B}\ }\textbf {\bibinfo {volume}
  {324}},\ \bibinfo {pages} {263} (\bibinfo {year} {1994})}\BibitemShut
  {NoStop}%
\end{thebibliography}%

\end{document}